\documentclass[iop]{emulateapj}

\def\actaa{Acta Astronomica}

\begin{document}

\shorttitle{PC and AC Relations for RR Lyrae Stars}
\shortauthors{Ngeow et al.}

\title{Period-Color and Amplitude-Color Relations at Maximum and Minimum Light for RR Lyrae Stars in the SDSS Stripe 82 Region}

\author{Chow-Choong Ngeow\altaffilmark{1}, Shashi M. Kanbur\altaffilmark{2}, Anupam Bhardwaj\altaffilmark{3,4}, Zachariah Schrecengost\altaffilmark{2} and Harinder P. Singh\altaffilmark{3}}
\altaffiltext{1}{Graduate Institute of Astronomy, National Central University, Jhongli 32001, Taiwan}
\altaffiltext{2}{Department of Physics, SUNY Oswego, Oswego, NY 13126, USA}
\altaffiltext{3}{Department of Physics and Astrophysics, University of Delhi, Delhi 110007, India}
\altaffiltext{4}{European Southern Observatory, Karl-Schwarzschild-Stra{\ss}e 2, Garching D-85748, Germany}

\begin{abstract}

Investigation of period-color (PC) and amplitude-color (AC) relations at the maximum and minimum light can be used to probe the interaction of the hydrogen ionization front (HIF) with photosphere and the radiation hydrodynamics of the outer envelopes of Cepheids and RR Lyraes. For example, theoretical calculations indicated that such interactions would occur at minimum light for RR Lyrae and result in a flatter PC relation. In the past, the PC and AC relations have been investigated by using either the $(V-R)_{MACHO}$ or $(V-I)$ colors. In this work, we extend previous work to other bands by analyzing the RR Lyraes in the Sloan Digital Sky Survey Stripe 82 Region. Multi-epoch data is available for RR Lyraes located within the footprint of the Stripe 82 Region in five ($ugriz$) bands. We present the PC and AC relations at maximum and minimum light in four colors: $(u-g)_0$, $(g-r)_0$, $(r-i)_0$ and $(i-z)_0$, after they are corrected for extinction. We found that the PC and AC relations for this sample of RR Lyraes show a complex nature in the form of flat, linear or quadratic relations. Furthermore, the PC relations at minimum light for fundamental mode RR Lyrae stars are separated according to the Oosterhoff type, especially in the $(g-r)_0$ and $(r-i)_0$ colors. If only considering the results from linear regressions, our results are quantitatively consistent with the theory of HIF-photosphere interaction for both fundamental and first overtone RR Lyraes.

\end{abstract}

\keywords{stars: variables: RR Lyrae --- stars: fundamental parameters --- stars: atmospheres}

\section{Introduction}\label{intro}

It has been well known that the helium II partial ionization zone drives the radial pulsation of Cepheids and RR Lyraes via the $\kappa$-mechanism. In between this zone and the photosphere, defined as a layer with an optical depth of $\tau= 2/3$, there exists another partial ionization zone -- the hydrogen ionization front (HIF). During the cycle of pulsation, the HIF will move ``in-and-out'' within the mass distribution. Furthermore, the relative positions of the HIF and stellar photosphere change with the pulsation phase. Hence, it is possible that HIF will interact with the photosphere at certain phases of the pulsation, where the photosphere is located at the base of the HIF. This has been demonstrated with a series of pulsation models constructed, for example, in \citet{simon1993}, \citet{kanbur1995}, \citet{kanbur1996}, and \citet{kanbur2004a}. Results based on these calculations suggested that this interaction will occur at maximum and and minimum light for Cepheids and RR Lyraes, respectively. 

The period-color (hereafter PC) and amplitude-color (hereafter AC) relations at maximum and minimum light can be used as diagnostics of such interactions for Cepheids and RR Lyraes. For instance, the interaction of HIF with the photosphere at certain pulsation phases would imply that the observed PC relation will be flat or very shallow (that is, the slope of the PC relation is close to zero) at these phases \citep[for a thorough discussion on this, see][and references therein]{bhardwaj2014}. In the case of Cepheids, the PC and AC relations have been extensively studied in a series of papers \citep{kanbur2004,kanbur2004a,kanbur2006,kanbur2007,kanbur2010}, as well as in \citet{bhardwaj2014}. For RR Lyraes, \citet{kanbur2005} investigated the PC and AC relations based on $4829$ RR Lyraes in Large Magellanic Cloud obtained from the MACHO project, and found that the PC and AC relations at minimum light have a slope close to zero for this set of RR Lyraes. A similar result was also found with the fundamental mode RR Lyraes in the Large and Small Magellanic Cloud using the Optical Gravitational Lensing Experiment III (OGLE-III) data \citep{bhardwaj2014}. Note that the flatness of the RR Lyraes PC relation at minimum light provides an alternate approach to estimate the reddening\citep{struch1966}.\footnote{Some examples of using the PC relation at minimum light to estimate reddening can be found in \citet{guldenschuh2005}, \citet{kunder2010}, and \citet{layden2013}. However, detailed investigation of such applications is beyond the scope of this paper.}

\begin{figure*}
  $\begin{array}{ccc}
    \includegraphics[angle=0,scale=0.29]{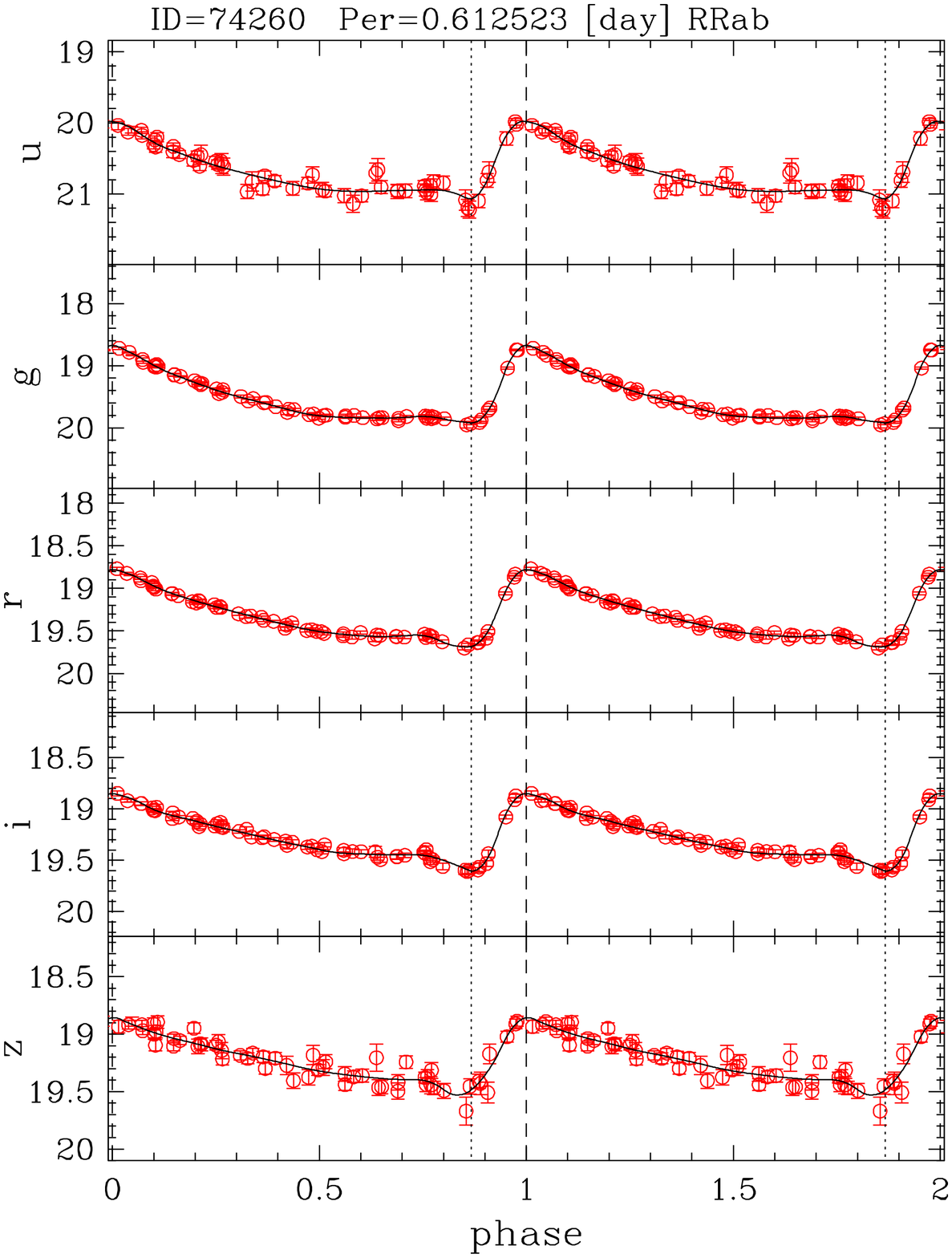} &
    \includegraphics[angle=0,scale=0.29]{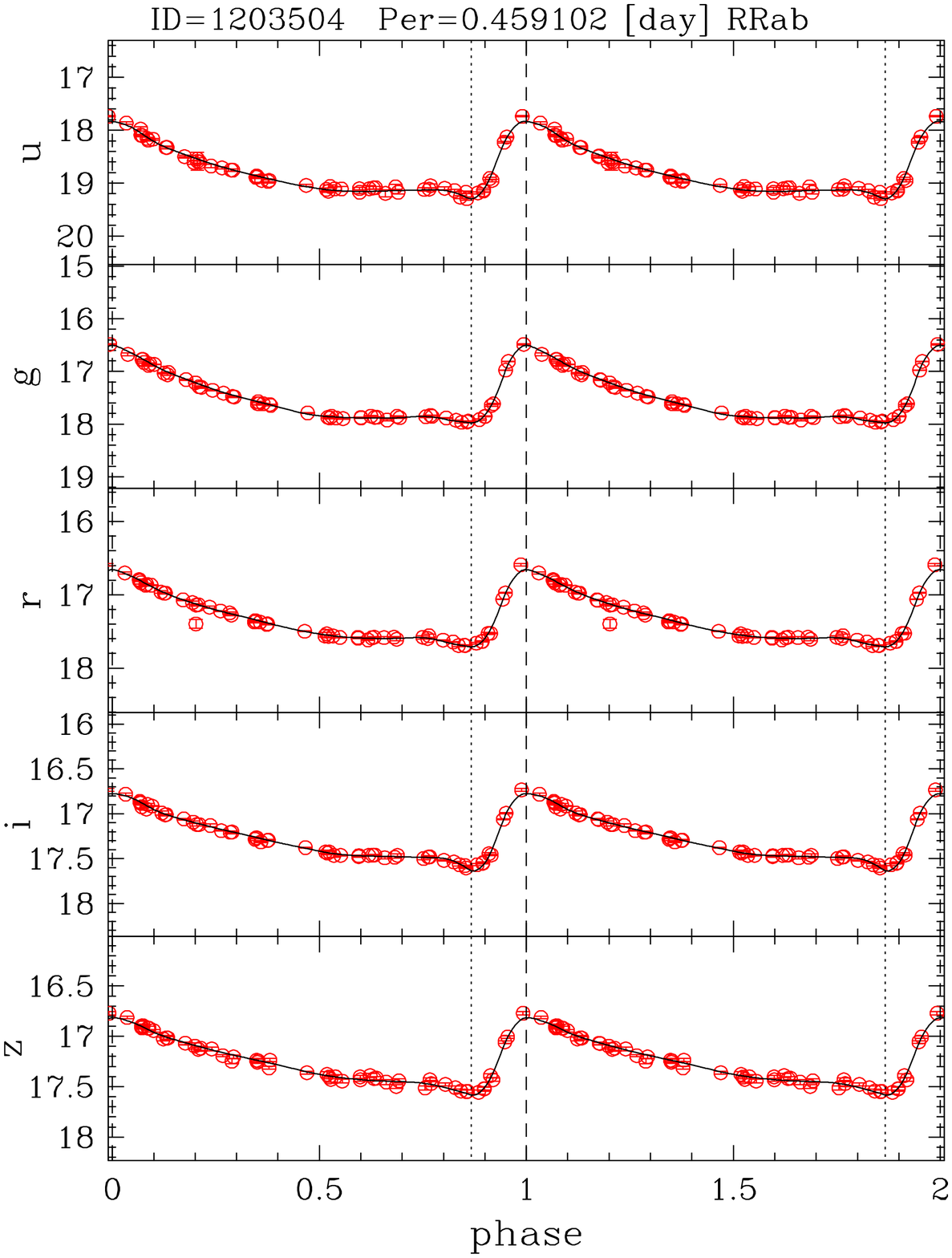} &
    \includegraphics[angle=0,scale=0.29]{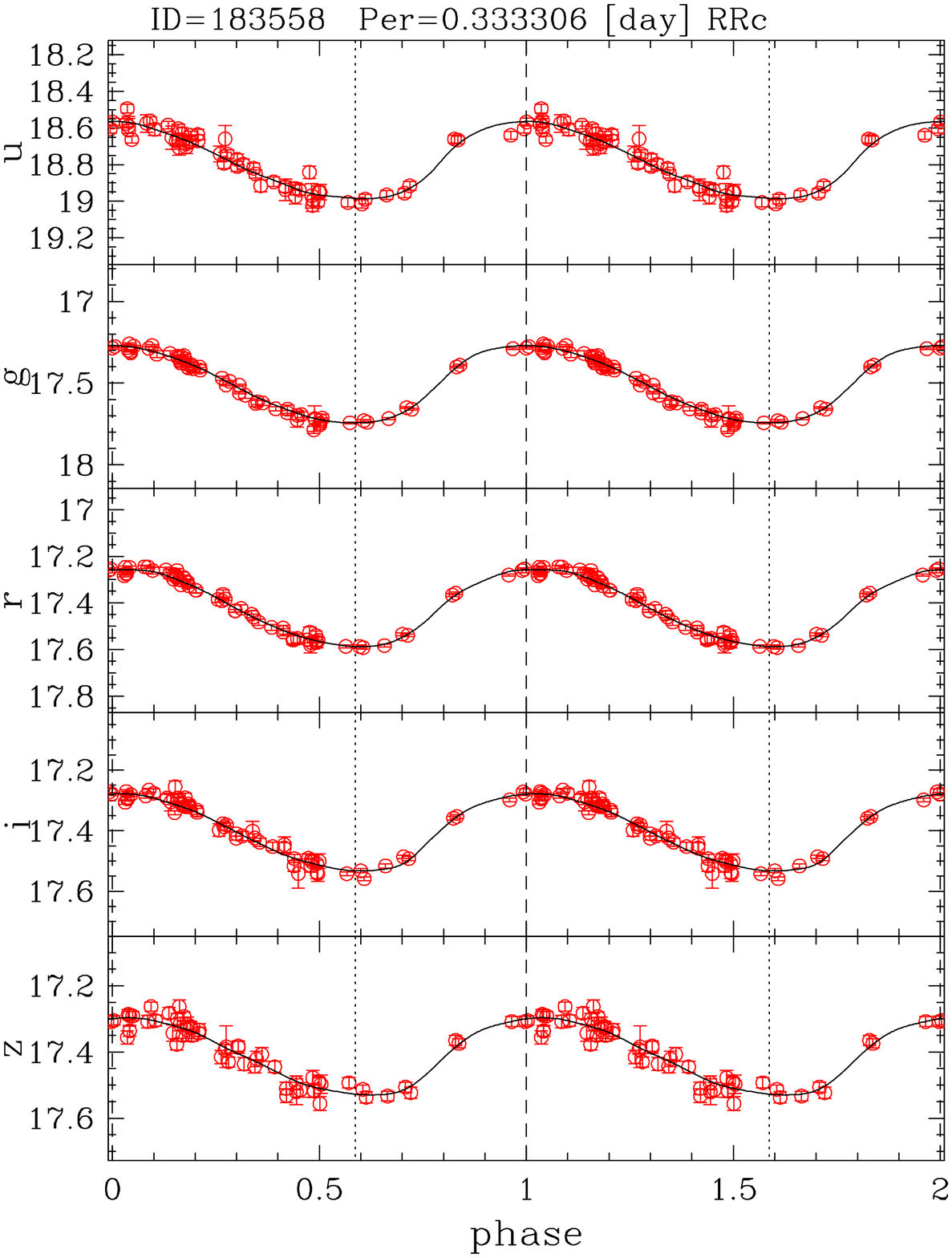} \\
  \end{array}$ 
  \caption{Examples of fitting the $ugriz$ template light curves to the light-curve data of RR Lyrae. Both of the light curves' data and the template light curves were taken from \citet{sesar2010}. The dashed and dotted vertical lines represent the phases at $g$-band maximum and minimum light, respectively. } \label{fig1}
\end{figure*}

In this paper, we extend the previous work to the RR Lyraes found in the Sloan Digital Sky Survey (SDSS) Stripe 82 region. \citet{kanbur2005} and \citet{bhardwaj2014} studied the PC and AC relations based on the $(V-R)_{MACHO}$ color and the $(V-I)$ color respectively. In contrast, the SDSS data enables our investigation of PC and AC relations in multi-colors. Preliminary results of this work can be found in \citet{bontorno2011}, and the main purpose of this paper is to present the final results based on the full analysis of the SDSS Stripe 82 RR Lyrae data. We first describe the data and methods used in this paper in Section \ref{data_method}. We then present the analysis of this set of data and the results in Section \ref{ana_result}. Finally, a discussion and conclusions of this paper are given in Section \ref{conclu}. 

\section{Data and Methods}\label{data_method}

The $ugriz$ band photometric light-curve data for RR Lyraes in the SDSS Stripe 82 region was adopted from \citet{sesar2010}, who found 483 RR Lyrae based on observations that spanned from 1998 September to 2007 November. Median numbers of data points per light curve are 55 and 56 in $uz$ bands and $gri$ bands, respectively. Further details regarding this dataset can be found in \citet{sesar2010}, and will not be repeated here. This set of RR Lyraes includes $379$ fundamental mode RRab stars and $104$ first overtone RRc stars. In addition to the light-curve data, \citet{sesar2010} also provided the $ugriz$ band template light curves for RR Lyraes. We adopted these template light curves to derive the amplitudes and colors at maximum and minimum light for each RR Lyrae in the sample as described further below.

In the first step, each of the light curves is converted to a phased light curve according to $\phi(t) = t/P - INT(t/P)$, where $P$ is the pulsation period as given in \citet{sesar2010} and the function $INT$ takes the integer part of the argument. The template light curves in a given band were then fitted to the phased light curves by determining the mean magnitudes, amplitudes, and the phase difference between the template and phased light curves. The \citet{nelder1965} simplex minimization algorithm, implemented within the {\tt GNU Scientific Library}, was employed to perform the fitting. 

\begin{figure}
  \plotone{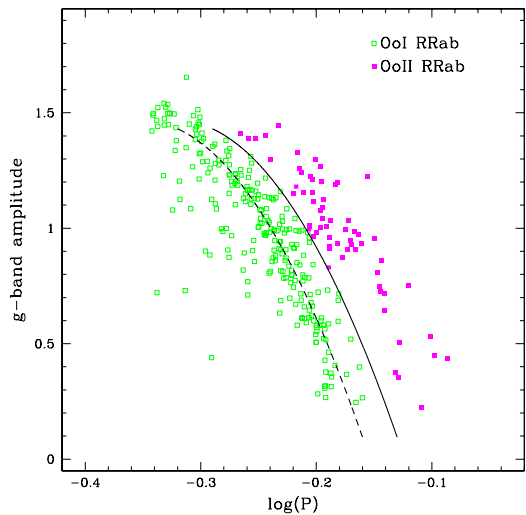}
  \caption{The period-amplitude diagram for RRab stars in our sample. The dashed curve is the quadratic regression fit to the locus of the data, as given in \citet[][their equation 21]{sesar2010}. Shifting this curve to the right by $0.03$ in $\log P$ yields the solid curve, which is the criterion given in \citet{sesar2010} to separate the OoI RRab (open green squares) and OoII RRab (filled magenta squares) stars. \label{fig_pamp}}
\end{figure}

\begin{deluxetable*}{llcc|cccc|cccc}
\tabletypesize{\scriptsize}
\tablecaption{Amplitudes and colors at the maximum and minimum light for RR Lyrae stars in the sample\label{tab1}}
\tablewidth{0pt} 
\tablehead{
\multicolumn{4}{c}{} &
\multicolumn{4}{c}{At Maximum Light} &
\multicolumn{4}{c}{At Minimum Light} 
\\
\colhead{ID\tablenotemark{a}} &
\colhead{Type} &
\colhead{$\log P$} &
\colhead{$AMP_g$\tablenotemark{b}} &
\colhead{$(u-g)_0$} &
\colhead{$(g-r)_0$} &
\colhead{$(r-i)_0$} &
\colhead{$(i-z)_0$} &
\colhead{$(u-g)_0$} &
\colhead{$(g-r)_0$} &
\colhead{$(r-i)_0$} &
\colhead{$(i-z)_0$} 
}
\startdata
4099	& ab-OoI & $-$0.19263 & 0.564 & 1.173 & 0.101 & 0.020 & 0.003 & 1.096 & 0.266 & 0.099 & 0.021 \\
13350	& ab-OoI & $-$0.26123 & 1.116 & 1.228 & $-$0.064 & $-$0.082 & $-$0.030 & 1.127 & 0.260 & 0.065 & 0.038 \\
15927   & ab-OoI & $-$0.21306 & 0.688 & 1.256 & 0.067 & $-$0.004 & $-$0.016 & 1.229 & 0.269 & 0.113 & 0.014 
\enddata
\tablecomments{The entire ASCII Table will be available in electronic form via the journal website. A portion of the Table is shown here for its layout and content.}
\tablenotetext{a}{ID from \citet{sesar2010}.}
\tablenotetext{b}{The $g$-band amplitude.}
\end{deluxetable*}

For a given RR Lyrae, the $g$-band best-fitted template light curve is used to determine the phases of maximum and minimum light, which are needed for constructing the PC and AC relations at these two extreme phases. We use this definition rather than the difference between the brightest and faintest observed points, in order to prevent amplitudes being effected by outliers. In other words, the SDSS colors $(\lambda_1-\lambda_2)$ at maximum light, where $\lambda_1<\lambda_2$, were determined from the same phase when the $g$ band is at maximum light. The same procedure was applied to determine the SDSS colors at minimum light. We emphasize that the colors at maximum light and minimum light were {\it not} based on the extreme values in a given light curve.\footnote{That is, the color $(\lambda_1-\lambda_2)$ at maximum light is not same as $(\lambda_1^{\mathrm{max}}-\lambda_2^{\mathrm{max}})$, where $\lambda^{\mathrm{max}}$ is the maximum light for light curve in bandpass $\lambda$.} The main reason to select the $g$-band light curve as the reference light curve for determining these two extreme phases is because at typical RR Lyrae effective temperatures (around $\sim 6000$K to $\sim 7000$K), Wien's displacement law suggests that the wavelengths at which the black-body curve peaks will fall within the $g$-band transmission curve. For the same reason amplitudes of these RR Lyraes will be based on the $g$-band fitted template light curves. Taking a reference band in this way and computing the colors at maximum and minimum light with respect to the reference band will preserve the phase differences in light curves at different wavebands. 

Figure \ref{fig1} presents examples of the fitted $ugriz$ band template light curves to randomly selected RR Lyraes in our sample. We visually examined all fitted light curves and removed those RR Lyraes that met either one of the following conditions: (a) light curves that show evidence of Blazhko or amplitude modulation,\footnote{Based on visual inspection, we suspect that $\sim40$ RRab and $\sim11$ RRc stars display evidence of Blazhko modulation. This is about 11\% of the sample studied here. However, a detailed study to confirm or falsify the Blazhko nature of these RR Lyraes is beyond the scope of this work.} and/or (b) light curves that do not exhibit well-defined colors at maximum and minimum light.\footnote{This could be due to, for example, sparse light curves that do not have data points around maximum and/or minimum light, light-curve data points that displayed large scatter, etc.} Our philosophy is governed by selecting a subset of well-fitted light curves, such as those presented in Figure \ref{fig1}, rather than including data that do not have well-determined colors at maximum and minimum light. After visual inspection, 312 and 86 RR Lyraes of $ab$- and $c$-types were left in our sample, respectively. Extinction corrections were determined using the $r$-band extinction values ($A_r$) given in \citet{sesar2010}; extinctions in other bands were scaled with the following relations:\footnote{Adopted from {\tt http://www.astro.washington.edu/users/\\ ivezic/sdss/catalogs/stripe82.html}} $A_u=1.873A_r$, $A_g=1.377A_r$, $A_i=0.758A_r$ and $A_z=0.537A_r$.

It is well known that the RRab stars in globular clusters exhibit the so-called Oosterhoff dichotomy \citep[for examples, see][]{catelan2009}: the Oosterhoff type I and II (hereafter OoI and OoII, respectively) can be distinguished via the period-amplitude (or Bailey) diagram \citep[for examples, see][]{clement2000,smith2011}. Before investigating the PC and AC relations for these RR Lyrae in the next section, we further divided our RRab stars into the OoI and OoII types by using the criterion suggested in \citet{sesar2010}. Figure \ref{fig_pamp} displays the period-amplitude diagram and the classified OoI ($N=248$) and OoII ($N=64$) RRab stars.

\section{Analysis and Results}\label{ana_result}

\begin{figure*} 
  \plottwo{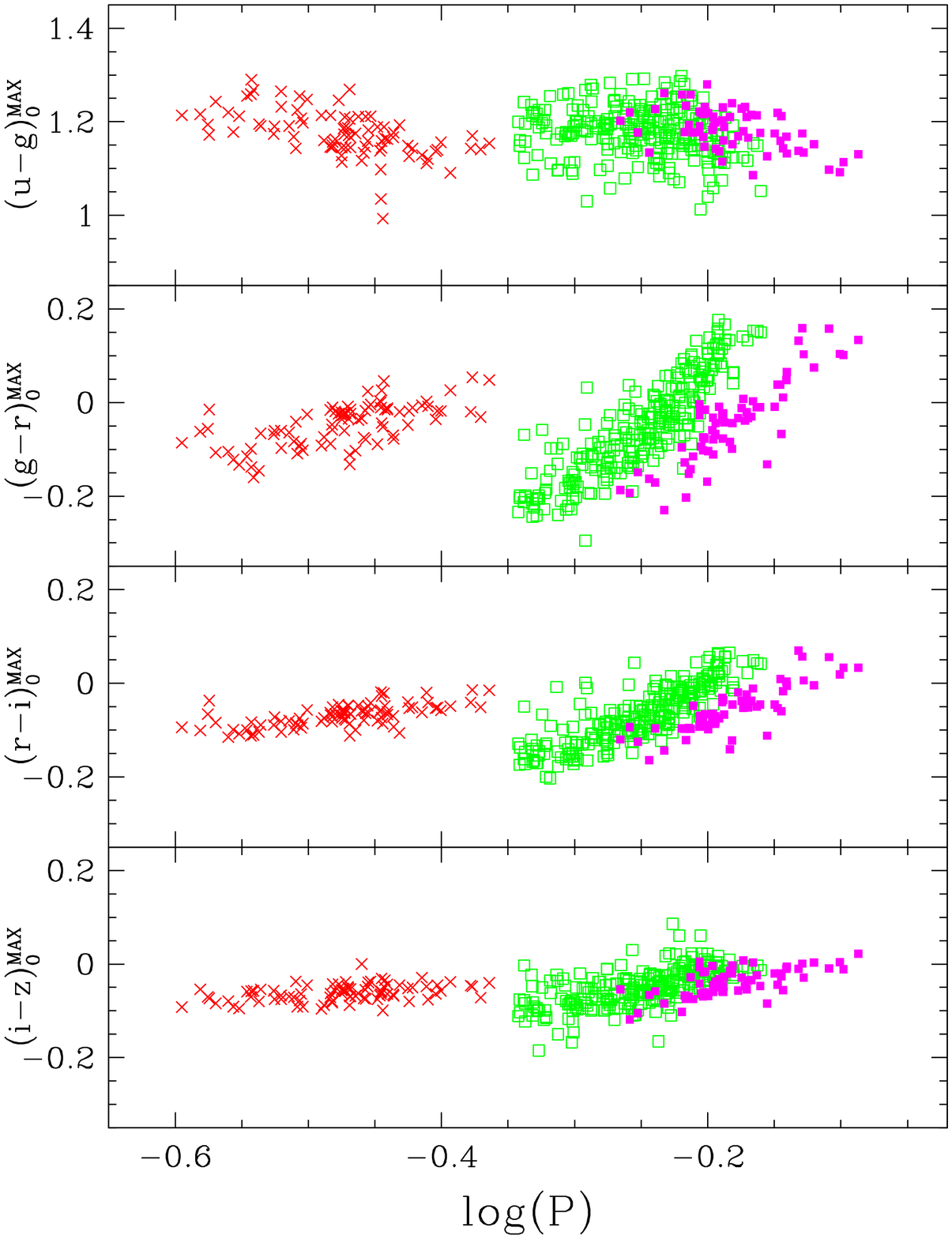}{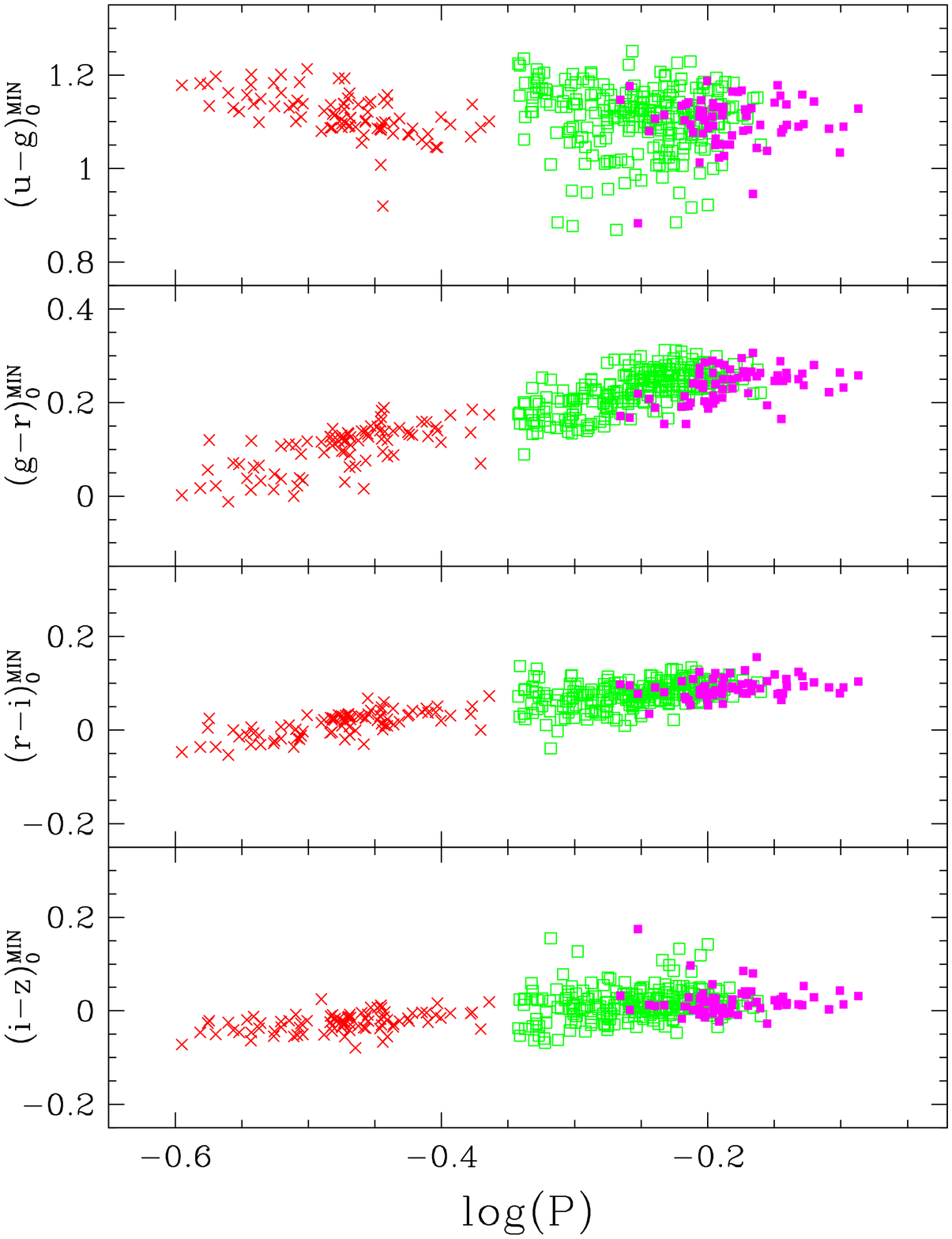}
  \caption{Period-color (PC) relations at ($g$-band) maximum (left panel) and minimum (right panel) light for the four extinction corrected SDSS colors. Green open squares and magenta filled squares represent the OoI RRab and OoII RRab stars, respectively, while the red crosses are for the RRc stars.} \label{fig_pcmaxmin}
\end{figure*}

\begin{figure*} 
  \plottwo{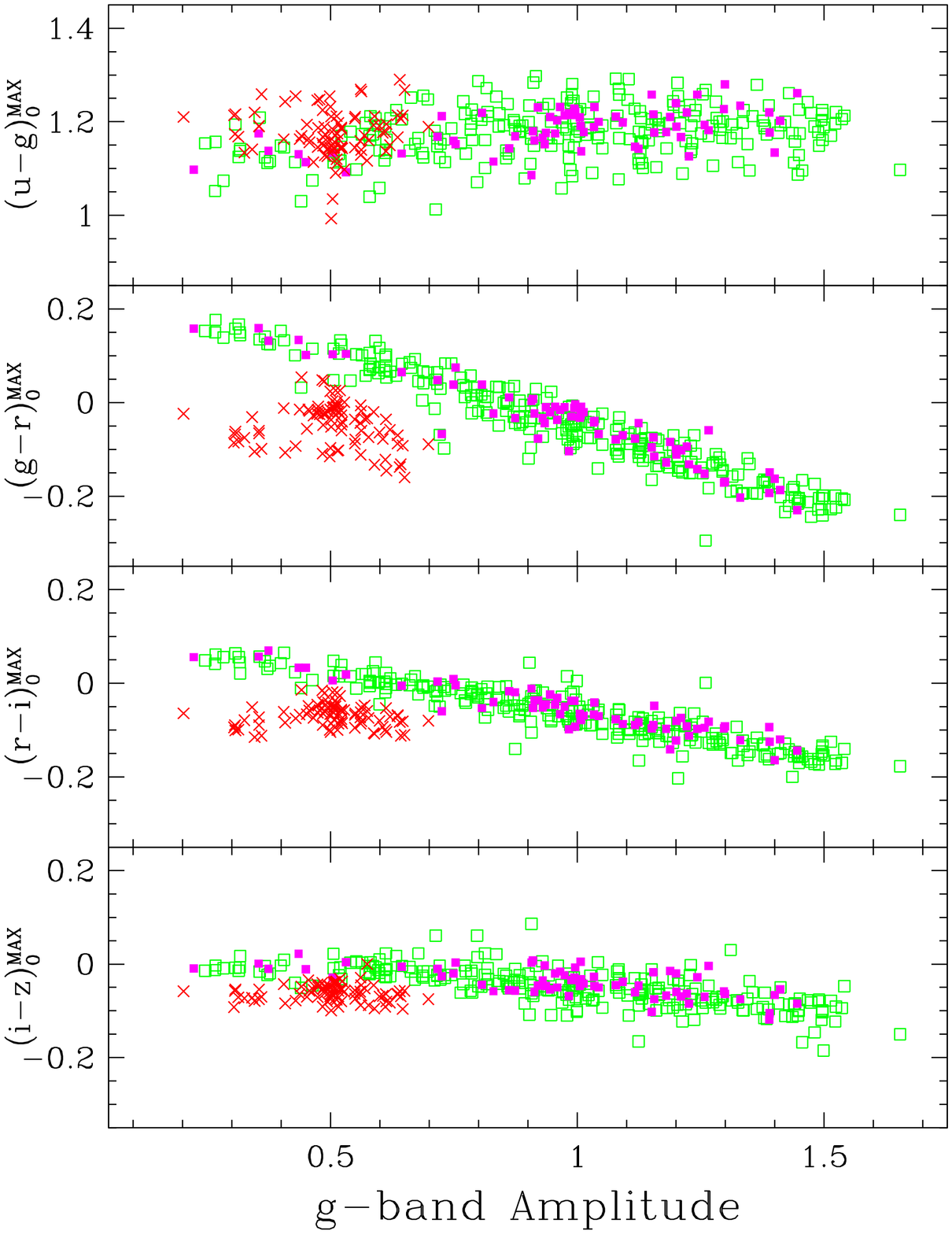}{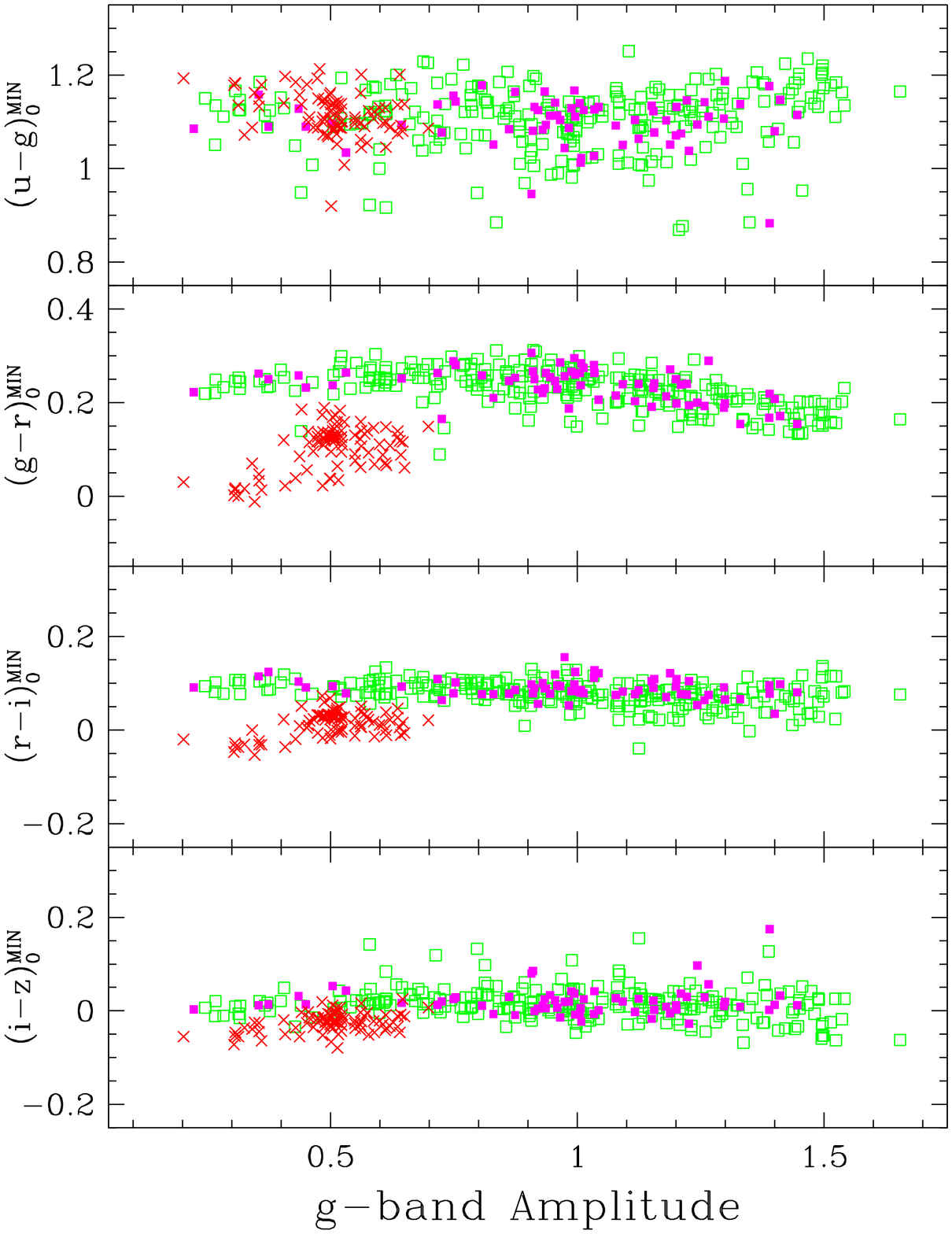}
  \caption{Amplitude-color (AC) relations at ($g$-band) maximum (left panel) and minimum (right panel) light for the four extinction corrected SDSS colors. Green open squares and magenta filled squares represent the OoI RRab and OoII RRab stars, respectively, while the red crosses are for the RRc stars.} \label{fig_acmaxmin}
\end{figure*}

\begin{figure*} 
  \plottwo{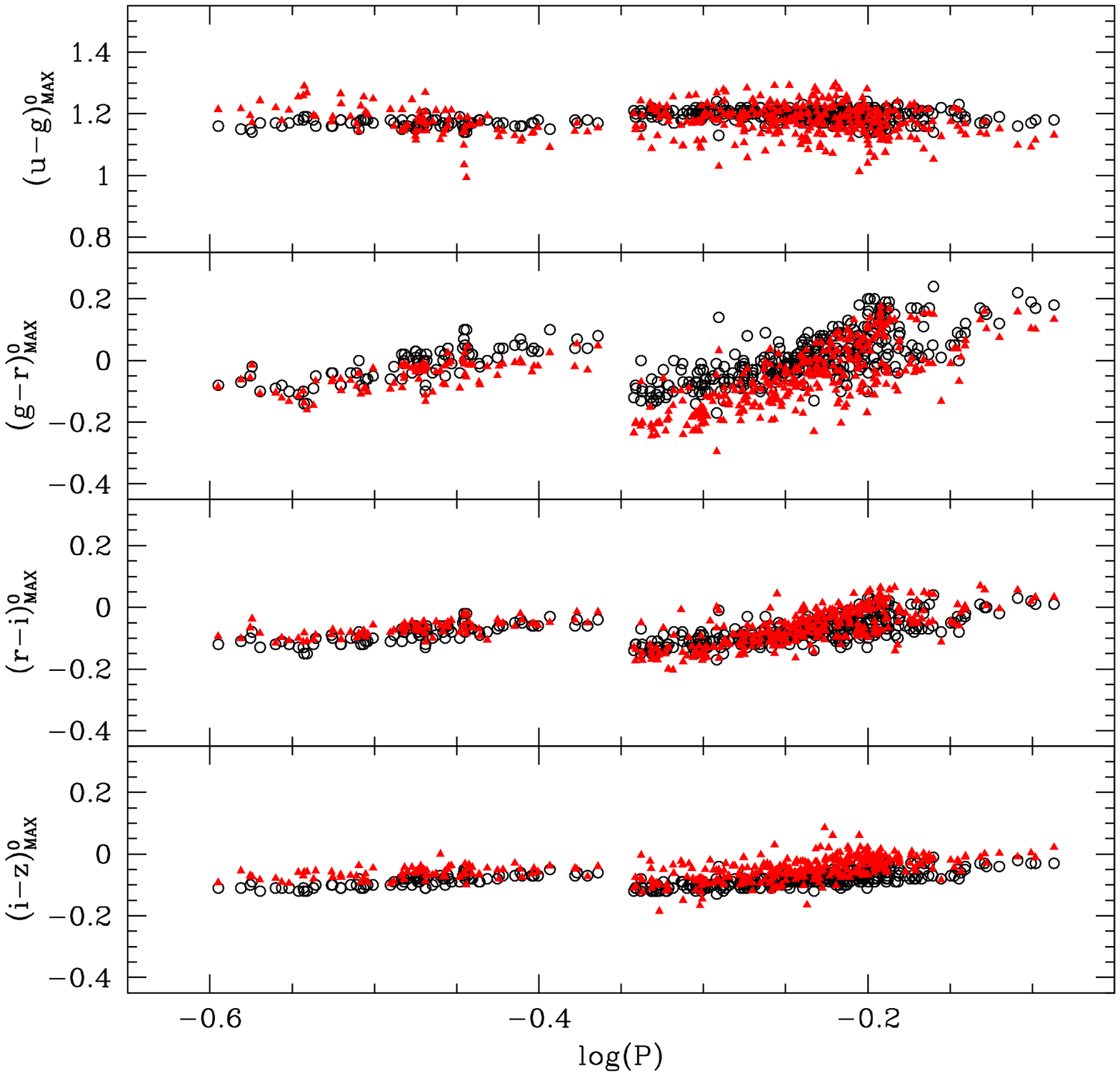}{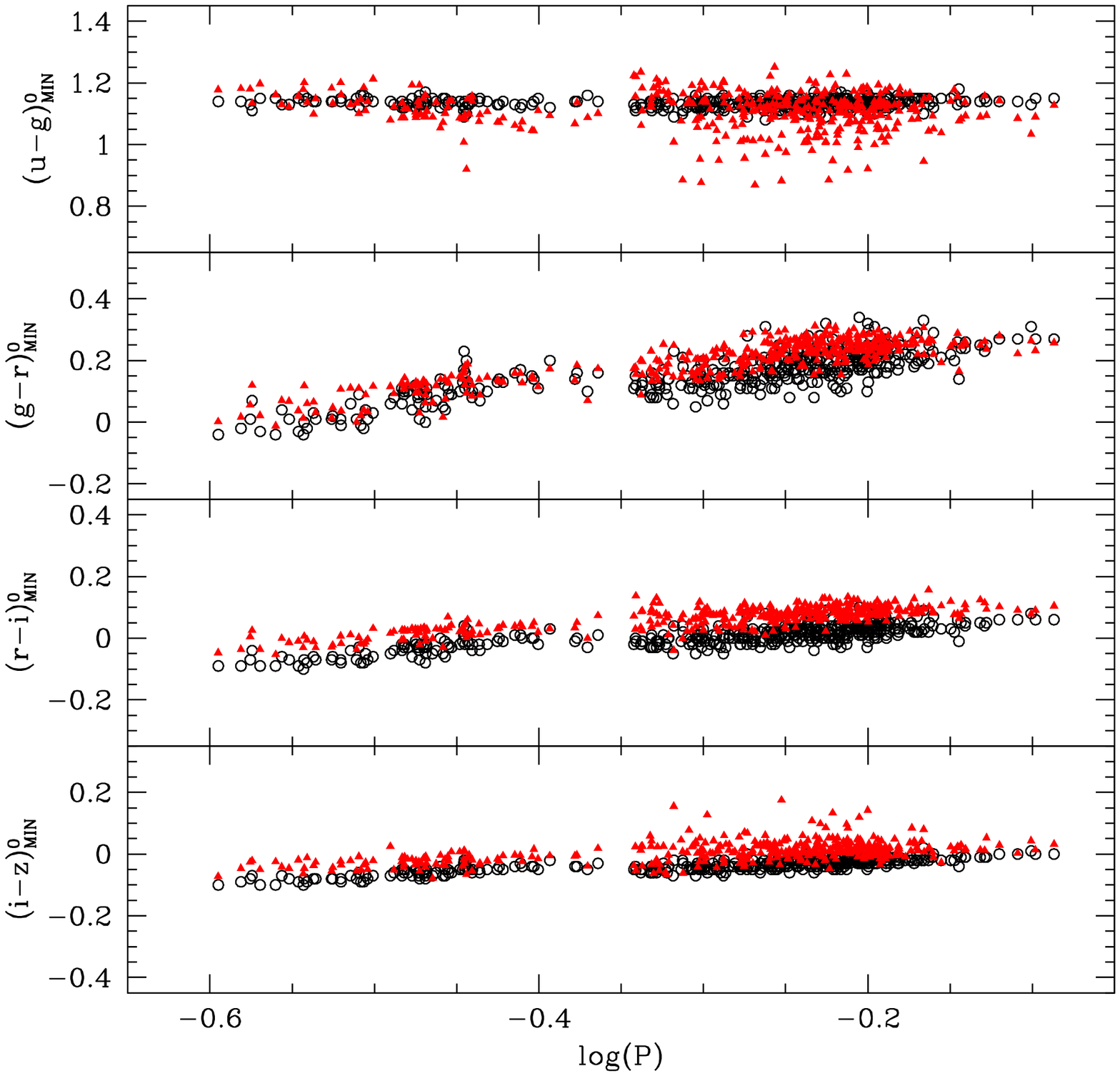}
  \caption{Comparison of the PC relations at ($g$-band) maximum (left panel) and minimum (right panel) light based on our results given in Table \ref{tab1} (red filled triangles) and the synthetic colors (black open circles) derived from synthetic magnitudes. The synthetic magnitudes in bandpass $\lambda$, $M_\lambda$, were calculated based on pseudo-color $C_0 = (u-g)_0 - (g-r)_0$ and the pulsation periods using the relations given in \citet[][their Figure 1]{caceres2008}. The synthetic colors are then derived as $M_{\lambda1}-M_{\lambda2}$, where $\lambda1 < \lambda2$. For RRc stars, we have fundamentalized the periods \citep[using the relation given in][]{catelan2009} before applying the conversion. For clarity, we did not separate the OoI RRab stars, OoII RRab stars, and RRc stars in this Figure.} \label{fig_vscc2008}
\end{figure*}

Table \ref{tab1} presents the $g$-band amplitudes and the colors at maximum and minimum light for our RR Lyrae sample. The corresponding plots of the PC and AC relations are displayed in Figure \ref{fig_pcmaxmin} and \ref{fig_acmaxmin}, respectively. We discuss our analysis and results on these relations further in the following subsections. The observed PC relations were also compared to the synthetic colors in Figure \ref{fig_vscc2008}.

\subsection{The PC Relations at Maximum Light}

\citet{kanbur2005} and \citet{bhardwaj2014} found a significant non-zero and positive slope for the PC relation at maximum light in the $(V-R)_{MACHO}$ and $(V-I)$ colors, respectively. In the left panel of Figure \ref{fig_pcmaxmin} we plotted the PC relations at maximum light for RRab and RRc stars: these display a variety of behavior. For example, we found significantly positive slopes for fundamental mode RRab stars for all colors, confirming the results presented in earlier work, except in $(u-g)_0$ color. The $(u-g)_0$ behavior at maximum light is more complicated. The RRab stars appeared to exhibit a flat relation with this becoming negative for the OoII RRab stars. In the case of first overtone RRc stars, shown as red crosses in Figure \ref{fig_pcmaxmin}, the $(u-g)_0$ PC slope is negative and a slight positive slope for all other colors. The negative PC slopes in the $(u-g)_0$ color and positive PC slopes in other longer wavelength colors could be understood due to the bolometric corrections when transforming the physical quantities (luminosity and effective temperature), to observable quantities such as magnitudes and colors \citep{caceres2008}. The left panel of Figure \ref{fig_vscc2008} compares the observed colors at maximum light, i.e. those from left panel of Figure \ref{fig_pcmaxmin}, to the synthetic colors derived using the relations given in \citet{caceres2008}, which demonstrates that the trends of the PC relations are very similar in both cases. 

Figure \ref{fig_pcmaxmin} also reveals that the PC relations at maximum light for RRab stars, especially in the $(g-r)_0$ and $(r-i)_0$ colors, can be split into two sequences that belong to OoI and OoII RRab stars, respectively (note that RRc stars do not split into two sequences in Figure \ref{fig_pcmaxmin}). Therefore, regressions on the PC relations were fitted separately to the OoI and OoII RRab stars. We first fit a linear regression to the PC relations at maximum light, and the fitted parameters are listed in columns (2) and (3) in Table \ref{tab_pc_maxmin}. We then applied a standard $t$-test to see if the slope ($\beta_1$) from the linear regression is consistent with zero or not, with null and alternate hypothesis as $H_0: \beta_1 = 0$ and $H_A: \beta_1 \neq 0$, respectively. By adopting a confidence level at $\alpha=0.05$, the null hypothesis can be rejected if the $p$-value, $p(t)$, is smaller than 0.05. The $t$-test results given in Table \ref{tab_pc_maxmin} show that $p(t)\sim0.00$ for all PC relations at maximum light. Hence, these PC relations exhibit a significant slope from the linear regression, regardless of the type of RR Lyrae (OoI RRab, OoII RRab, or RRc). In addition, the linear slopes are different in each SDSS color between the OoI RRab, OoII RRab, and RRc stars. 

Inspecting the left panel of Figure \ref{fig_pcmaxmin} suggests that the PC relation at maximum light for RRab stars might not be linear, especially in the $(u-g)_0$, $(g-r)_0$, and $(r-i)_0$ colors. We therefore fitted a quadratic regression to the PC relation at maximum light; the fitted parameters are summarized in columns (7)-(9) in Table \ref{tab_pc_maxmin}. Dispersions ($\sigma$) from the fitted regressions, listed in columns (4) and (10) of Table \ref{fig_pcmaxmin}, might not be a good metric to decide whether linear or quadratic regressions are better fits to the data, as the dispersions show marginal or no improvement when moving from linear to quadratic regressions. Instead, the standard $F$-test should be applied to evaluate which regression method is better. We calculated the $F$-value as follows.

\begin{eqnarray}
F & = & \frac{(RSS_l-RSS_q)/[(N-2)-(N-3)]}{RSS_q/(N-3)}, \nonumber
\end{eqnarray}

\noindent where $N$ is number of data points, and $RSS_l$ and $RSS_q$ are the residual sums of squares for the linear and quadratic regressions, respectively. We also evaluated the $p$-value under the $F$-distribution $F_{1,N-3}$ with the null hypothesis that linear regression is sufficient to fit the data. The alternate hypothesis is that the quadratic regression is needed. Results of the $F$-test are given in columns (11) and (12) of Table \ref{fig_pcmaxmin}. As before, we adopted $\alpha=0.05$ such that the null hypothesis can be rejected if $p(F)<0.05$. 

In the case of RRc stars, the $F$-test provides evidence that a linear regression can be used to fit the PC relation at maximum light in the four SDSS colors. For RRab stars, the $F$-test suggested that a quadratic regression is a better model for OoI RRab star in the $(u-g)_0$, $(g-r)_0$, and $(r-i)_0$ colors. The $(i-z)_0$ color, on the other hand, shows a marginal $F$-test result. This behavior is different from the OoII RRab stars because the PC relation at maximum light in $(u-g)_0$ color is (marginally) nonlinear but linear in the other three colors.

\begin{deluxetable*}{lrrcccrrrccc}
\tabletypesize{\scriptsize}
%\rotate
\tablecaption{Period-color relations at maximum and minimum light using linear and quadratic regression \label{tab_pc_maxmin}}
\tablewidth{0pt}
\tablehead{
\colhead{} & \multicolumn{3}{c}{Linear Regression} & \colhead{} & \colhead{} & \multicolumn{4}{c}{Quadratic Regression} & \colhead{} & \colhead{} \\
\colhead{Color} &
\colhead{$\beta_0$} &
\colhead{$\beta_1$} &
\colhead{$\sigma$} &
\colhead{$t$} & 
\colhead{$p(t)$} &
\colhead{$\beta_0$} &
\colhead{$\beta_1$} &
\colhead{$\beta_2$} &
\colhead{$\sigma$} & 
\colhead{$F$} &
\colhead{$p(F)$}  \\
\colhead{(1)} &
\colhead{(2)} &
\colhead{(3)} &
\colhead{(4)} &
\colhead{(5)} &
\colhead{(6)} &
\colhead{(7)} &
\colhead{(8)} &
\colhead{(9)} &
\colhead{(10)} &
\colhead{(11)} &
\colhead{(12)}
}
\startdata
%\cutinhead{At Maximum Light} 
\hline
\multicolumn{12}{c}{At Maximum Light} \\
\hline
\multicolumn{12}{c}{OoI RRab} \\
$(u-g)_0$ & $1.122\pm0.019$ & $-0.222\pm0.076$ & 0.052 & 2.94 & 0.00 & $0.782\pm0.104$ & $-2.961\pm0.825$ & $-5.343\pm1.602$ & 0.051 & 11.1 & 0.00 \\
$(g-r)_0$ & $0.476\pm0.019$ & $2.106\pm0.074$ & 0.051 & 28.5 & 0.00 & $0.978\pm0.099$ & $6.145\pm0.782$ & $7.878\pm1.520$ & 0.048 & 26.9 & 0.00 \\
$(r-i)_0$ & $0.232\pm0.013$ & $1.187\pm0.050$ & 0.034 & 23.7 & 0.00 & $0.506\pm0.068$ & $3.395\pm0.540$ & $4.306\pm1.048$ & 0.033 & 16.9 & 0.00 \\
$(i-z)_0$ & $0.096\pm0.012$ & $0.580\pm0.046$ & 0.032 & 12.6 & 0.00 & $0.207\pm0.064$ & $1.472\pm0.509$ & $1.741\pm0.988$ & 0.031 & 3.10 & 0.08 \\
\multicolumn{12}{c}{OoII RRab} \\ 
$(u-g)_0$ & $1.076\pm0.023$ & $-0.596\pm0.122$ & 0.037 & 4.87 & 0.00 & $0.939\pm0.070$ & $-2.238\pm0.806$ & $-4.672\pm2.269$ & 0.036 & 4.24 & 0.04 \\
$(g-r)_0$ & $0.326\pm0.027$ & $2.029\pm0.145$ & 0.044 & 14.0 & 0.00 & $0.355\pm0.086$ & $2.378\pm0.989$ & $0.991\pm2.784$ & 0.045 & 0.13 & 0.72 \\
$(r-i)_0$ & $0.139\pm0.017$ & $1.081\pm0.093$ & 0.028 & 11.6 & 0.00 & $0.199\pm0.055$ & $1.794\pm0.627$ & $2.029\pm1.765$ & 0.028 & 1.32 & 0.25 \\
$(i-z)_0$ & $0.054\pm0.014$ & $0.512\pm0.075$ & 0.023 & 6.82 & 0.00 & $0.023\pm0.044$ & $0.142\pm0.509$ & $-1.052\pm1.431$ & 0.023 & 0.54 & 0.47 \\
\multicolumn{12}{c}{RRc} \\ 
$(u-g)_0$ & $0.921\pm0.042$ & $-0.536\pm0.087$ & 0.041 & 6.13 & 0.00 & $0.973\pm0.306$ & $-0.317\pm1.282$ & $0.228\pm1.333$ & 0.041 & 0.03 & 0.86 \\
$(g-r)_0$ & $0.229\pm0.036$ & $0.582\pm0.075$ & 0.035 & 7.74 & 0.00 & $0.427\pm0.263$ & $1.414\pm1.100$ & $0.867\pm1.143$ & 0.036 & 0.58 & 0.45 \\
$(r-i)_0$ & $0.073\pm0.019$ & $0.298\pm0.039$ & 0.019 & 7.57 & 0.00 & $0.258\pm0.137$ & $1.077\pm0.572$ & $0.811\pm0.594$ & 0.018 & 1.86 & 0.18 \\
$(i-z)_0$ & $0.005\pm0.017$ & $0.139\pm0.035$ & 0.016 & 4.01 & 0.00 & $-0.053\pm0.122$ & $-0.102\pm0.509$ & $-0.252\pm0.529$ & 0.016 & 0.23 & 0.64 \\
%\cutinhead{At Minimum Light} 
\hline
\multicolumn{12}{c}{At Minimum Light} \\
\hline
\multicolumn{12}{c}{OoI RRab} \\
$(u-g)_0$ & $1.062\pm0.025$ & $-0.183\pm0.099$ & 0.068 & 1.85 & 0.07 & $\cdots$ & $\cdots$ & $\cdots$ & $\cdots$ & $\cdots$ & $\cdots$ \\
$(g-r)_0$ & $0.384\pm0.012$ & $0.625\pm0.046$ & 0.031 & 13.7 & 0.00 & $0.155\pm0.062$ & $-1.220\pm0.495$ & $-3.598\pm0.962$ & 0.031 & 14.0 & 0.00 \\
$(r-i)_0$ & $0.140\pm0.009$ & $0.248\pm0.035$ & 0.024 & 7.11 & 0.00 & $0.216\pm0.049$ & $0.858\pm0.388$ & $1.190\pm0.754$ & 0.024 & 2.49 & 0.12 \\
$(i-z)_0$ & $0.050\pm0.012$ & $0.139\pm0.047$ & 0.032 & 2.94 & 0.00 & $-0.051\pm0.066$ & $-0.666\pm0.525$ & $-1.570\pm1.020$ & 0.032 & 2.37 & 0.13 \\
\multicolumn{12}{c}{OoII RRab} \\ 
$(u-g)_0$ & $1.108\pm0.032$ & $0.043\pm0.175$ & 0.053 & 0.24 & 0.81 & $\cdots$ & $\cdots$ & $\cdots$ & $\cdots$ & $\cdots$ & $\cdots$ \\
$(g-r)_0$ & $0.314\pm0.020$ & $0.430\pm0.106$ & 0.032 & 4.05 & 0.00 & $0.147\pm0.059$ & $-1.573\pm0.676$ & $-5.701\pm1.902$ & 0.030 & 8.99 & 0.00 \\
$(r-i)_0$ & $0.116\pm0.012$ & $0.147\pm0.066$ & 0.020 & 2.24 & 0.03 & $0.100\pm0.039$ & $-0.043\pm0.446$ & $-0.540\pm1.255$ & 0.020 & 0.19 & 0.67 \\
$(i-z)_0$ & $0.014\pm0.019$ & $-0.035\pm0.101$ & 0.031 & 0.34 & 0.73 & $\cdots$ & $\cdots$ & $\cdots$ & $\cdots$ & $\cdots$ & $\cdots$ \\
\multicolumn{12}{c}{RRc} \\ 
$(u-g)_0$ & $0.866\pm0.039$ & $-0.528\pm0.082$ & 0.038 & 6.46 & 0.00 & $1.130\pm0.285$ & $0.587\pm1.193$ & $1.162\pm1.240$ & 0.039 & 0.88 & 0.35 \\
$(g-r)_0$ & $0.408\pm0.036$ & $0.644\pm0.076$ & 0.036 & 8.47 & 0.00 & $0.053\pm0.263$ & $-0.855\pm1.102$ & $-1.561\pm1.146$ & 0.036 & 1.86 & 0.18 \\
$(r-i)_0$ & $0.186\pm0.019$ & $0.364\pm0.041$ & 0.019 & 8.96 & 0.00 & $0.024\pm0.141$ & $-0.321\pm0.591$ & $-0.714\pm0.614$ & 0.019 & 1.35 & 0.25 \\
$(i-z)_0$ & $0.063\pm0.018$ & $0.185\pm0.039$ & 0.018 & 4.79 & 0.00 & $0.107\pm0.135$ & $0.370\pm0.567$ & $0.192\pm0.589$ & 0.018 & 0.11 & 0.75 
\enddata
\tablecomments{The linear regression takes the form of $C=\beta_0 + \beta_1 \log P$, where $C$ denotes the four (extinction corrected) SDSS colors. Similarly, the quadratic regression takes the form of $C=\beta_0 + \beta_1 \log P + \beta_2 [\log P]^2$. $\sigma$ is the dispersion from the regression fits. Results from the $t$-test, the $t$-values, and the corresponding $p$-values, $p(t)$, are listed in columns (5) and (6). The $t$-test only tests for the slopes ($\beta_1$) in linear regression if they are consistent with zero or not. On the other hand, the $F$-test was applied to test whether or not quadratic regression is a better model to describe the data. The $F$-test results are given in columns (11) and (12). The quadratic regression and the $F$-test were only applied to those relations showing non-zero slopes from the $t$-test.}
\end{deluxetable*}

\begin{deluxetable*}{lrrcccrrrccc}
\tabletypesize{\scriptsize}
%\rotate
\tablecaption{Amplitude-color relations at maximum and minimum light using linear and quadratic regression \label{tab_ac_maxmin}}
\tablewidth{0pt}
\tablehead{
\colhead{} & \multicolumn{3}{c}{Linear Regression} & \colhead{} & \colhead{} & \multicolumn{4}{c}{Quadratic Regression} & \colhead{} & \colhead{} \\
\colhead{Color} &
\colhead{$\beta_0$} &
\colhead{$\beta_1$} &
\colhead{$\sigma$} &
\colhead{$t$} & 
\colhead{$p(t)$} &
\colhead{$\beta_0$} &
\colhead{$\beta_1$} &
\colhead{$\beta_2$} &
\colhead{$\sigma$} & 
\colhead{$F$} &
\colhead{$p(F)$}  \\
\colhead{(1)} &
\colhead{(2)} &
\colhead{(3)} &
\colhead{(4)} &
\colhead{(5)} &
\colhead{(6)} &
\colhead{(7)} &
\colhead{(8)} &
\colhead{(9)} &
\colhead{(10)} &
\colhead{(11)} &
\colhead{(12)} 
}
\startdata
%\cutinhead{At Maximum Light} \\
\hline
\multicolumn{12}{c}{At Maximum Light} \\
\hline
\multicolumn{12}{c}{OoI RRab} \\
$(u-g)_0$ & $1.135\pm0.011$ & $0.043\pm0.010$ & 0.051 & 4.25 & 0.00 & $1.060\pm0.024$ & $0.223\pm0.052$ & $-0.095\pm0.027$ & 0.050 & 12.2 & 0.00 \\
$(g-r)_0$ & $0.260\pm0.006$ & $-0.315\pm0.006$ & 0.031 & 51.4 & 0.00 & $0.241\pm0.015$ & $-0.270\pm0.032$ & $-0.024\pm0.017$ & 0.031 & 1.97 & 0.16 \\
$(r-i)_0$ & $0.112\pm0.005$ & $-0.180\pm0.005$ & 0.024 & 37.6 & 0.00 & $0.096\pm0.011$ & $-0.142\pm0.025$ & $-0.020\pm0.013$ & 0.024 & 2.31 & 0.13 \\
$(i-z)_0$ & $0.038\pm0.006$ & $-0.088\pm0.006$ & 0.029 & 15.3 & 0.00 & $0.000\pm0.014$ & $0.002\pm0.030$ & $-0.048\pm0.016$ & 0.028 & 9.39 & 0.00 \\
\multicolumn{12}{c}{OoII RRab} \\
$(u-g)_0$ & $1.098\pm0.017$ & $0.087\pm0.017$ & 0.037 & 5.19 & 0.00 & $1.072\pm0.037$ & $0.155\pm0.084$ & $-0.038\pm0.046$ & 0.037 & 0.68 & 0.41 \\
$(g-r)_0$ & $0.261\pm0.014$ & $-0.307\pm0.013$ & 0.029 & 22.8 & 0.00 & $0.204\pm0.029$ & $-0.164\pm0.065$ & $-0.081\pm0.036$ & 0.029 & 5.02 & 0.03 \\
$(r-i)_0$ & $0.109\pm0.009$ & $-0.168\pm0.009$ & 0.019 & 18.9 & 0.00 & $0.106\pm0.020$ & $-0.159\pm0.045$ & $-0.005\pm0.025$ & 0.020 & 0.04 & 0.84 \\
$(i-z)_0$ & $0.030\pm0.011$ & $-0.070\pm0.011$ & 0.023 & 6.57 & 0.00 & $-0.003\pm0.023$ & $0.013\pm0.052$ & $-0.047\pm0.029$ & 0.023 & 2.61 & 0.11 \\
\multicolumn{12}{c}{RRc} \\
$(u-g)_0$ & $1.176\pm0.030$ & $-0.002\pm0.059$ & 0.050 & 0.03 & 0.98 & $\cdots$ & $\cdots$ & $\cdots$ & $\cdots$ & $\cdots$ & $\cdots$ \\
$(g-r)_0$ & $-0.001\pm0.027$ & $-0.093\pm0.054$ & 0.046 & 1.72 & 0.09 & $\cdots$ & $\cdots$ & $\cdots$ & $\cdots$ & $\cdots$ & $\cdots$ \\
$(r-i)_0$ & $-0.066\pm0.014$ & $-0.006\pm0.028$ & 0.024 & 0.22 & 0.83 & $\cdots$ & $\cdots$ & $\cdots$ & $\cdots$ & $\cdots$ & $\cdots$ \\
$(i-z)_0$ & $-0.060\pm0.011$ & $-0.004\pm0.021$ & 0.018 & 0.18 & 0.86 & $\cdots$ & $\cdots$ & $\cdots$ & $\cdots$ & $\cdots$ & $\cdots$ \\
%\cutinhead{At Minimum Light} \\
\hline
\multicolumn{12}{c}{At Minimum Light} \\
\hline
\multicolumn{12}{c}{OoI RRab} \\
$(u-g)_0$ & $1.086\pm0.014$ & $0.022\pm0.014$ & 0.068 & 1.60 & 0.11 & $\cdots$ & $\cdots$ & $\cdots$ & $\cdots$ & $\cdots$ & $\cdots$ \\
$(g-r)_0$ & $0.295\pm0.007$ & $-0.068\pm0.007$ & 0.035 & 9.62 & 0.00 & $0.201\pm0.016$ & $0.158\pm0.034$ & $-0.119\pm0.018$ & 0.033 & 44.8 & 0.00 \\
$(r-i)_0$ & $0.107\pm0.005$ & $-0.029\pm0.005$ & 0.025 & 5.94 & 0.00 & $0.119\pm0.012$ & $-0.059\pm0.026$ & $0.016\pm0.013$ & 0.025 & 1.38 & 0.24 \\
$(i-z)_0$ & $0.034\pm0.007$ & $-0.020\pm0.006$ & 0.032 & 3.08 & 0.00 & $-0.015\pm0.015$ & $0.099\pm0.033$ & $-0.063\pm0.017$ & 0.032 & 13.2 & 0.00 \\
\multicolumn{12}{c}{OoII RRab} \\
$(u-g)_0$ & $1.108\pm0.025$ & $-0.007\pm0.024$ & 0.053 & 0.30 & 0.76 & $\cdots$ & $\cdots$ & $\cdots$ & $\cdots$ & $\cdots$ & $\cdots$ \\
$(g-r)_0$ & $0.289\pm0.016$ & $-0.054\pm0.015$ & 0.033 & 3.57 & 0.00 & $0.180\pm0.030$ & $0.223\pm0.067$ & $-0.156\pm0.037$ & 0.029 & 17.6 & 0.00 \\
$(r-i)_0$ & $0.108\pm0.009$ & $-0.018\pm0.009$ & 0.020 & 1.97 & 0.05 & $\cdots$ & $\cdots$ & $\cdots$ & $\cdots$ & $\cdots$ & $\cdots$ \\
$(i-z)_0$ & $0.012\pm0.014$ & $0.009\pm0.014$ & 0.031 & 0.63 & 0.53 & $\cdots$ & $\cdots$ & $\cdots$ & $\cdots$ & $\cdots$ & $\cdots$ \\
\multicolumn{12}{c}{RRc} \\
$(u-g)_0$ & $1.206\pm0.027$ & $-0.179\pm0.052$ & 0.044 & 3.43 & 0.00 & $1.314\pm0.084$ & $-0.657\pm0.357$ & $0.509\pm0.377$ & 0.044 & 1.82 & 0.18 \\
$(g-r)_0$ & $-0.032\pm0.025$ & $0.268\pm0.050$ & 0.042 & 5.41 & 0.00 & $-0.300\pm0.074$ & $1.463\pm0.317$ & $-1.272\pm0.334$ & 0.039 & 14.5 & 0.00 \\
$(r-i)_0$ & $-0.052\pm0.014$ & $0.131\pm0.028$ & 0.024 & 4.65 & 0.00 & $-0.220\pm0.041$ & $0.879\pm0.177$ & $-0.797\pm0.186$ & 0.022 & 18.3 & 0.00 \\
$(i-z)_0$ & $-0.069\pm0.011$ & $0.087\pm0.022$ & 0.019 & 3.88 & 0.00 & $-0.119\pm0.036$ & $0.310\pm0.153$ & $-0.237\pm0.161$ & 0.019 & 2.17 & 0.14 
\enddata
\tablecomments{The linear regression takes the form of $C=\beta_0 + \beta_1 AMP_g$, where $C$ denotes the four (extinction corrected) SDSS colors and $AMP_g$ is the $g$-band amplitude. Similarly, the quadratic regression takes the form of $C=\beta_0 + \beta_1 AMP_g + \beta_2 [AMP_g]^2$. $\sigma$ is the dispersion from the regression fits. Results from the $t$-test, the $t$-values, and the corresponding $p$-values, $p(t)$, are listed in columns (5) and (6). The $t$-test only tests for the slopes ($\beta_1$) in linear regression whether or not they are consistent with zero. On the other hand, the $F$-test was applied to test whether or not quadratic regression is a better model to describe the data. The $F$-test results are given in columns (11) and (12). The quadratic regressions and the $F$-test were only applied to those relations showing non-zero slopes from the $t$-test.} 
\end{deluxetable*}

\subsection{The PC Relations at Minimum Light}

\begin{figure*}
  $\begin{array}{cc}
    \includegraphics[angle=0,scale=0.41]{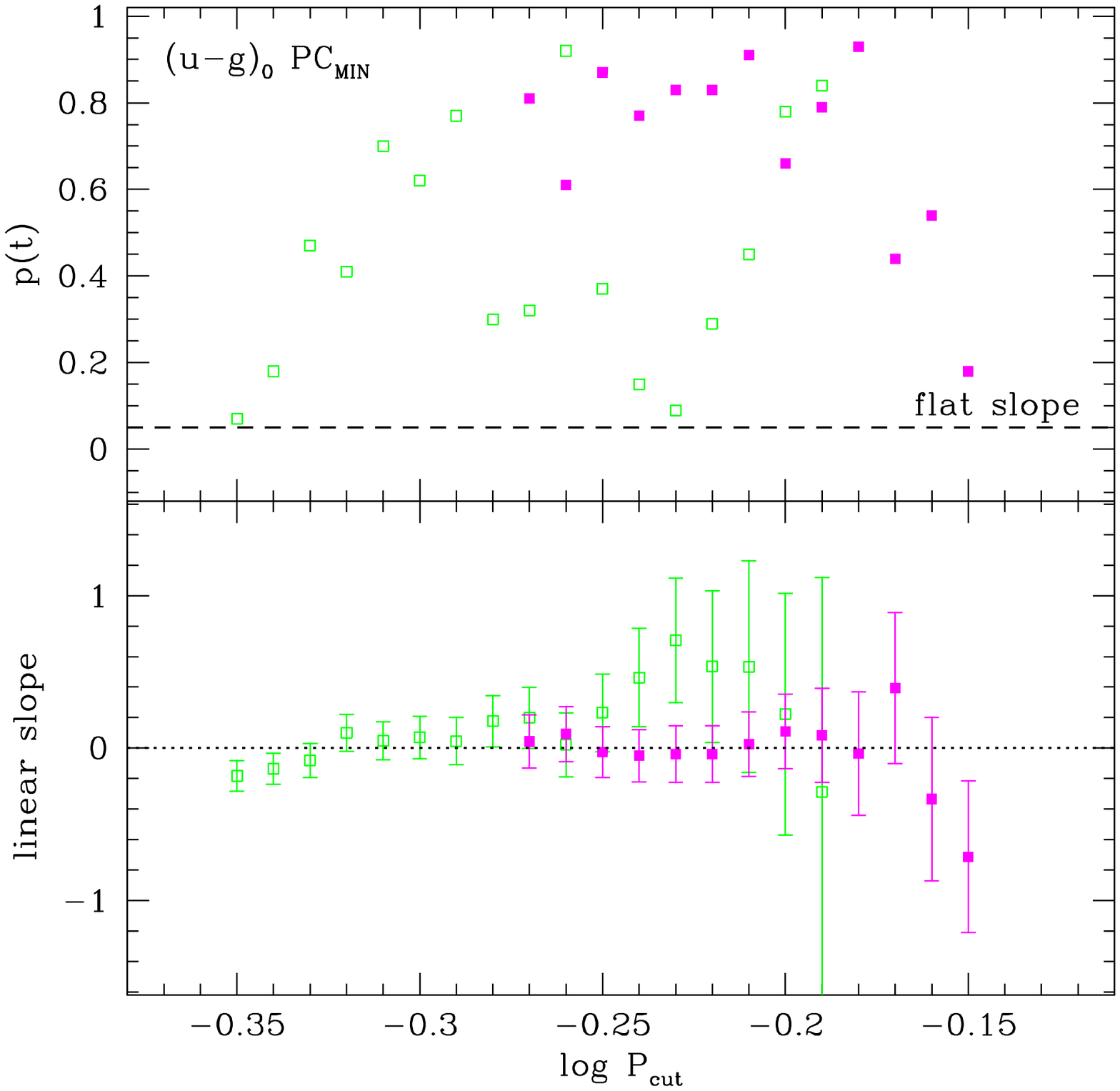} & 
    \includegraphics[angle=0,scale=0.41]{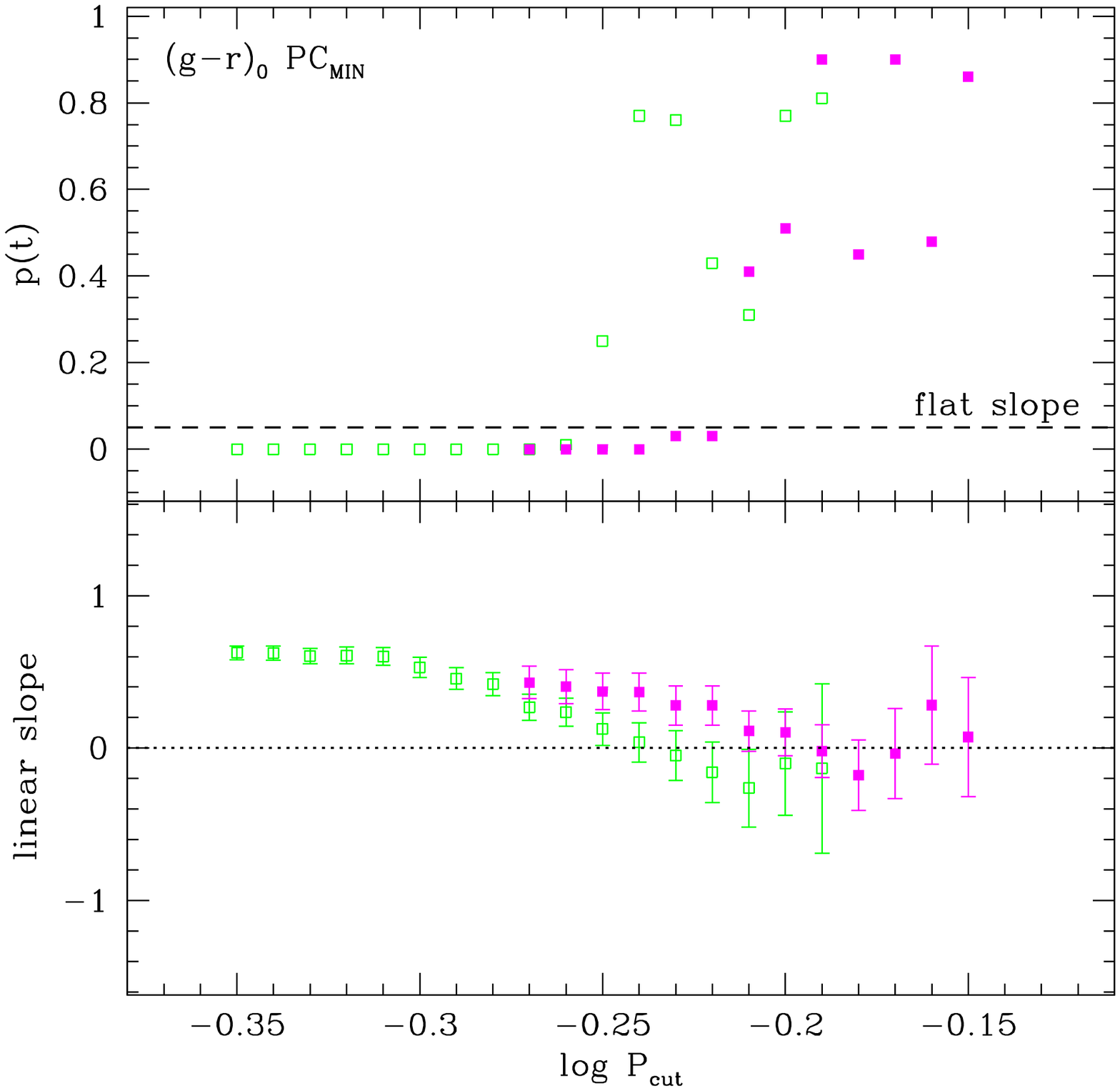} \\
    \includegraphics[angle=0,scale=0.41]{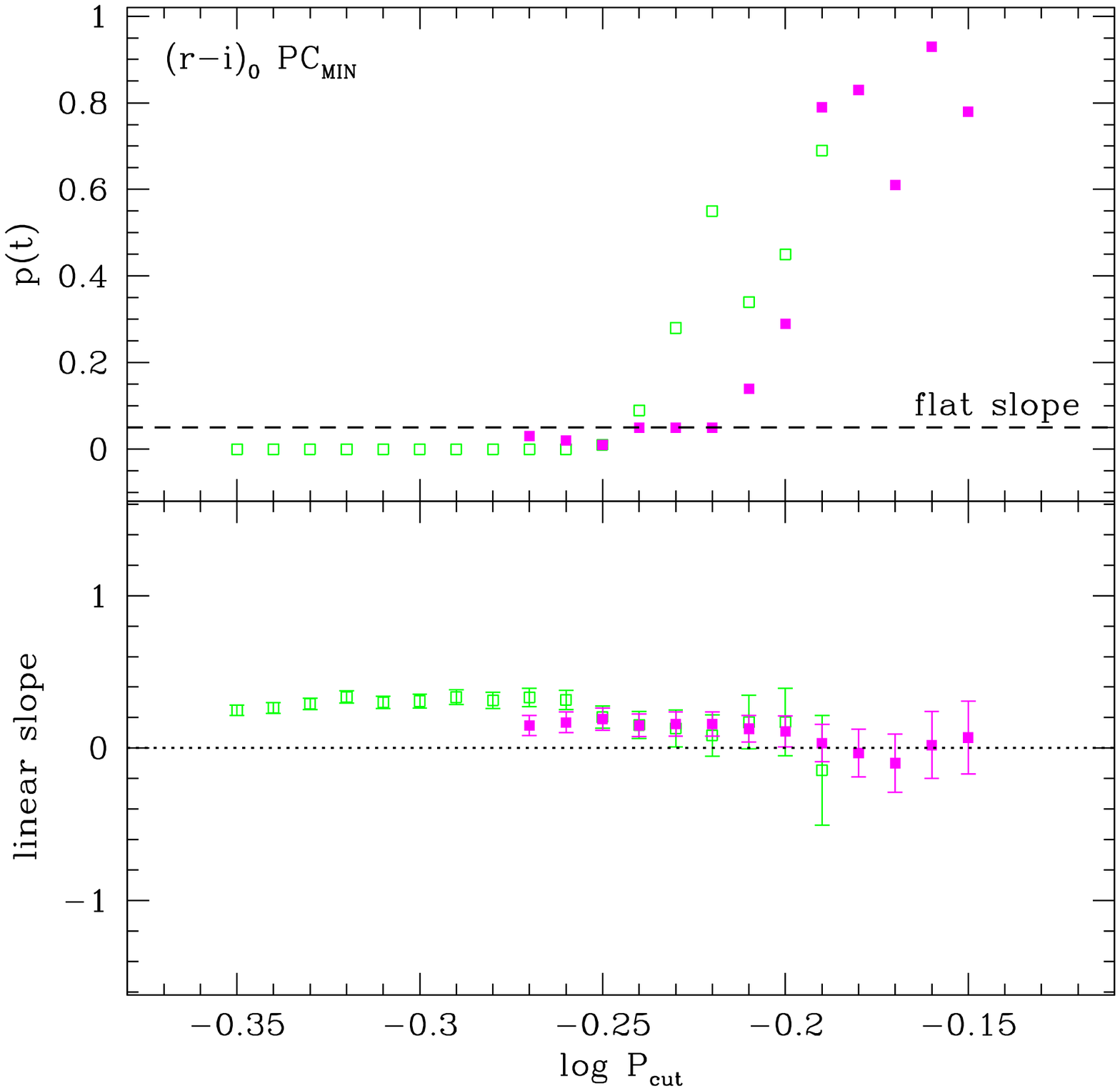} & 
    \includegraphics[angle=0,scale=0.41]{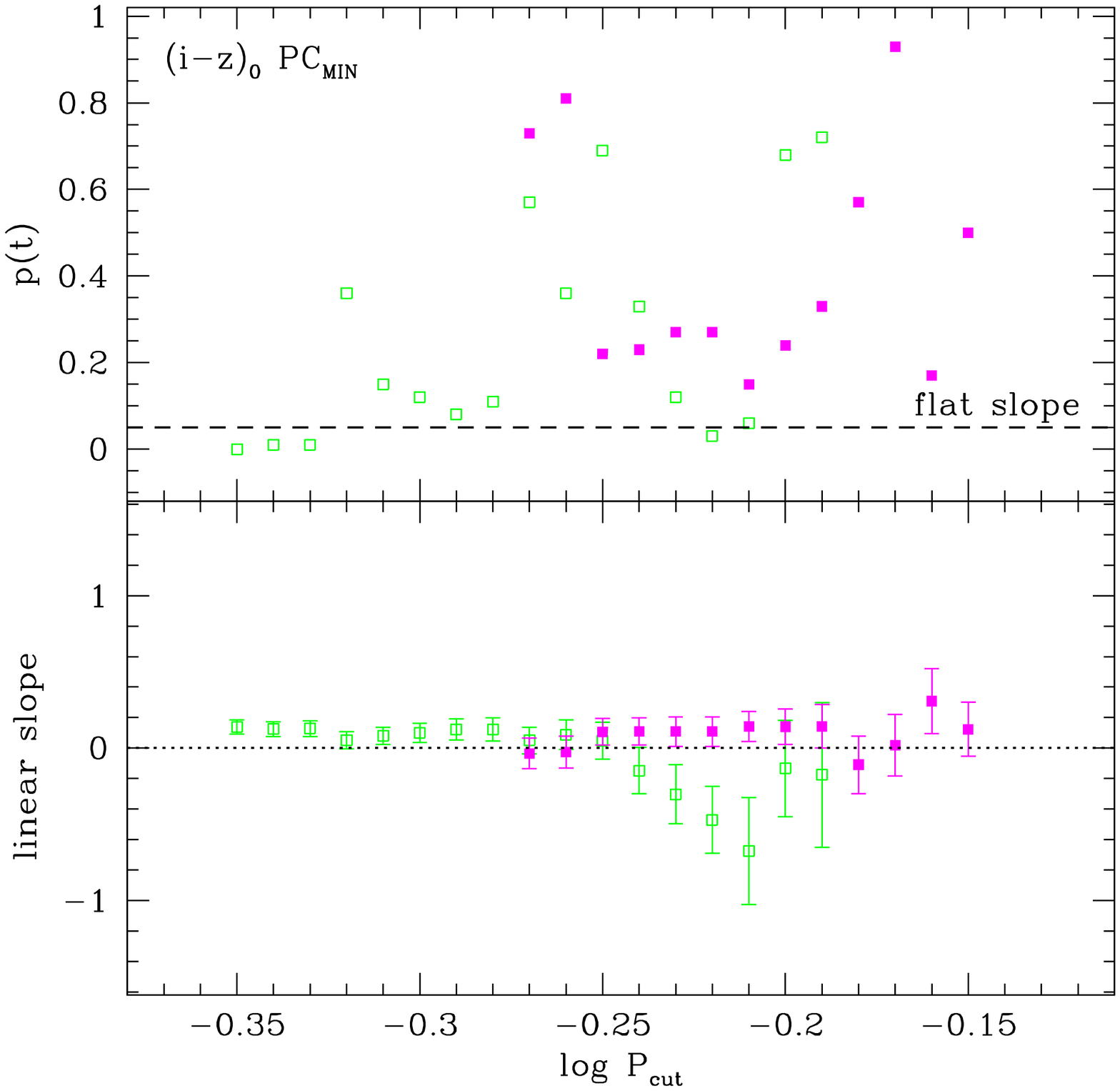} \\    
  \end{array}$ 
  \caption{Results of the $t$-test on PC slopes at minimum light (upper panels in each sub-Figures) and the slopes of linear regression (lower panels in each sub-Figures) as a function of $\log P_{\mathrm{cut}}$, where the linear regressions were fitted to data with $\log P > \log P_{\mathrm{cut}}$. We stopped the fitting of linear regression when the number of data is below 10 after removing those RRab stars with $\log P < \log P_{\mathrm{cut}}$. The dashed lines represent the adopted confidence level of $0.05$: slopes of the regressions are considered flat if $p(t)>0.05$. The dotted lines indicate the zero slopes, and not the fitting of the data points. Green open squares and magenta filled squares represent the OoI RRab and OoII RRab stars, respectively.} \label{fig_pccut}
\end{figure*}

Previous theoretical work, as described in the Introduction, suggested that the PC relation for RR Lyrae is almost independent of pulsation periods at minimum light. In the case of RRab stars, this result was verified from the MACHO data presented in \citet{kanbur2005} and the OGLE-III data as described in \citet{bhardwaj2014}. The RRc stars do not exhibit a flat PC relation at minimum light for the reasons discussed in \citet{bhardwaj2014}. The right panel of Figure \ref{fig_pcmaxmin} presents the PC relations at minimum light in the four SDSS colors. From this Figure, clear non-zero slopes of the PC relations at minimum light were found in all four SDSS colors for the RRc stars: this is also seen in the Magellanic Cloud RRc stars in the $(V-I)$ color \citep{bhardwaj2014}. As in previous subsections, we fitted the linear and quadratic regressions, as well as applied $t$- and $F$-tests to these PC relations. Our results presented in Table \ref{tab_pc_maxmin} confirm the existence of a linear PC relation at minimum light for these type of pulsating stars. 

In the case of RRab stars, the majority of the PC slopes at minimum light, as presented in Table \ref{tab_pc_maxmin}, do not show a clear evidence of flatness, which is different to what has been found from previous works. However, we note that in almost all cases except one, the PC slope at minimum light is significantly shallower than the PC slope at maximum light. The exception is the $(u-g)_0$ color for OoI RRab stars. For this dataset, combining the $t$-test and $F$-test results, the PC relations at minimum light for OoI RRab stars are better described with a quadratic regression in the $(g-r)_0$ color, but linear in the $(r-i)_0$ and $(i-z)_0$ colors. The $t$-test result of the $(u-g)_0$ PC relation at minimum light of OoI RRab stars suggests a marginal flat relation. The OoII RRab stars, however, display a flat PC relation at minimum light in the $(u-g)_0$ and $(i-z)_0$ colors, but a quadratic and linear relation in the $(g-r)_0$ and $(r-i)_0$ colors, respectively. Finally, we note that trends of the observed PC relations at minimum light are very similar to the PC relation constructed from synthetic colors (right panel of Figure \ref{fig_vscc2008}).

At first glance, our results on PC relations at minimum light seem to disagree with theoretical expectations \citep{kanbur1996} and earlier empirical studies \citep{kanbur2005,bhardwaj2014}. A careful inspection on the right panel of Figure \ref{fig_pcmaxmin}, especially the $(g-r)_0$ PC relation, reveals that a flat PC relation could exist if we consider only RRab stars with $\log P$ greater than approximately $-0.2$. Therefore, we applied a period cut at $\log P_{\mathrm{cut}}$ and fitted a linear regression only for those RRab stars with periods greater than this period cut. This was done separately for OoI and OoII RRab stars. The slopes of the linear regression as a function of the period cut are displayed in lower panels of each sub-panel in Figure \ref{fig_pccut}. We also applied the $t$-test on these slopes, and the corresponding $p(t)$ values are shown in the upper panels of the sub-Figures in the same Figure. Our results show that $(u-g)_0$ PC relations at minimum light are always flat regardless of period range and Oosterhoff type. The same result is also found for $(i-z)_0$ OoII PC relations at minimum light. For other PC relations at minimum light, a flat relation exists only if we consider RRab stars with periods greater than a given $\log P_{\mathrm{cut}}$. For example, we observed a flat $(g-r)_0$ PC relation at minimum light for OoI RRab stars with $\log P > -0.25$.  

\subsection{The AC Relations at Maximum and Minimum Light}

\begin{figure*} 
  \plottwo{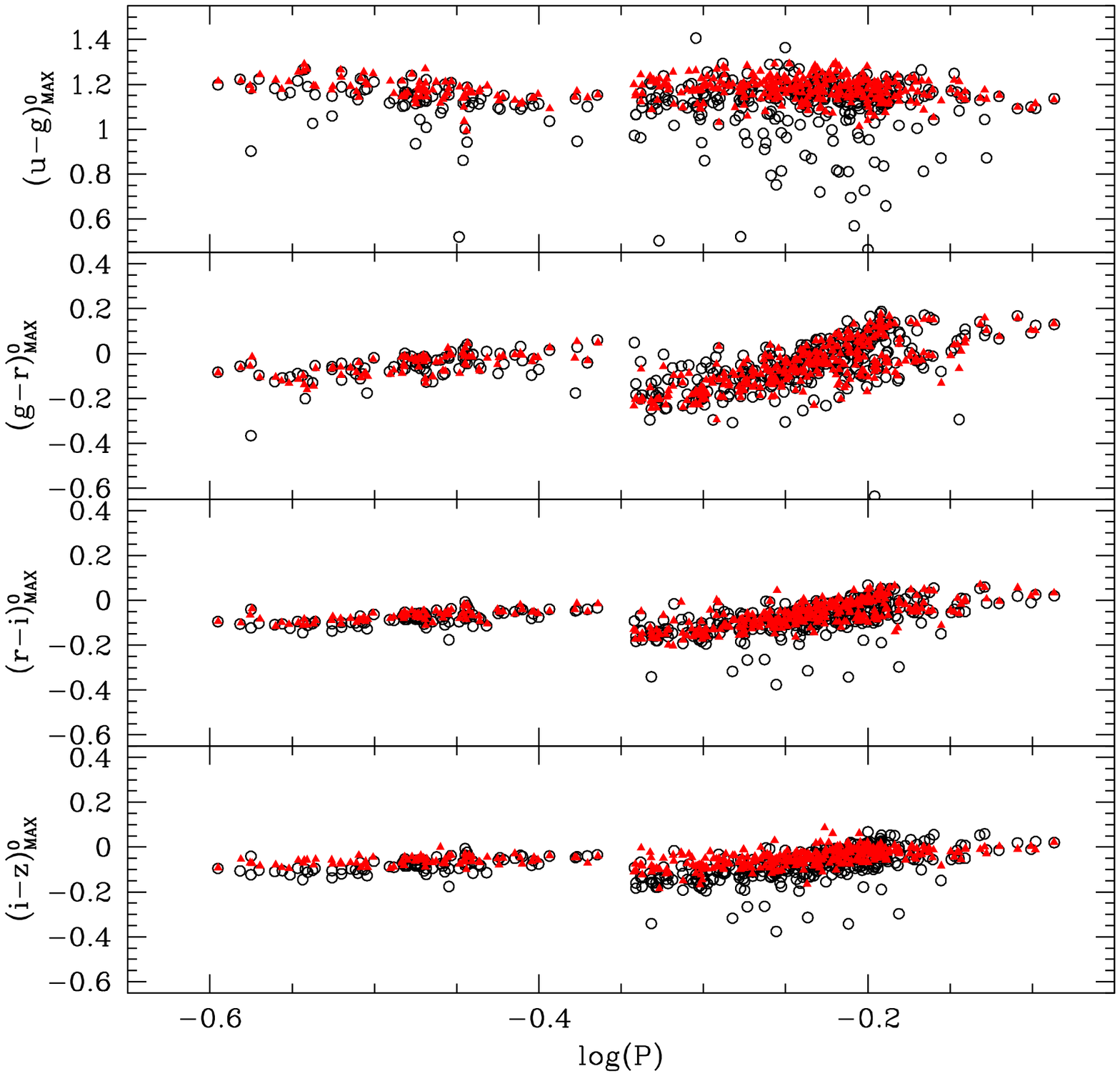}{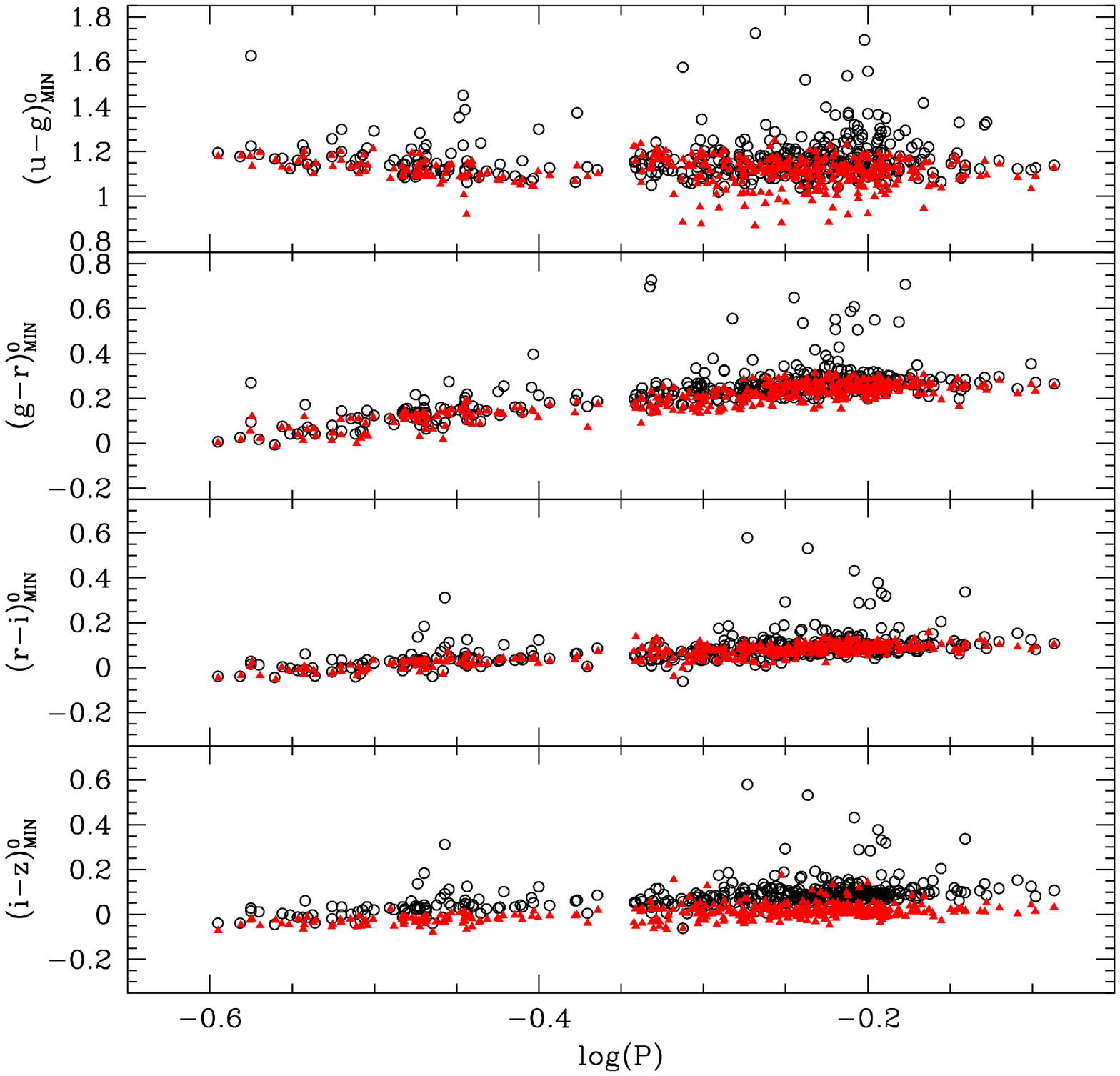}
  \caption{Comparison of the PC relations at ($g$-band) maximum (left panel) and minimum (right panel) light based on the template light-curve fitting approach (red filled triangles) and Fourier decomposition technique (black open circles) to determine the colors. We used a sixth-order Fourier expansion \citep[an example of such an expression can be found, for example, in][]{deb2009} to fit the $ugriz$-band light-curve data. The problem of larger scatter for the data points based on the Fourier decomposition technique could be potentially remedied using different fitting orders in the Fourier decomposition technique, for example, by applying Baart's condition \citep{deb2009} to provide a better fit to individual light curves. For clarity, we did not separate the OoI RRab stars, OoII RRab stars, and RRc stars in this Figure.} \label{fig_vsfourier}
\end{figure*}

As in the case of PC relations, the AC relations presented in Figure \ref{fig_acmaxmin} were fitted with linear and quadratic regressions and the results are summarized in Table \ref{tab_ac_maxmin}. For RRc stars, flat AC relations are found at maximum light in all four colors, even though the $(g-r)_0$ AC relation shows a marginal result. We note that \citet{bhardwaj2014} also found a significantly different AC relation at maximum light in the $(V-I)$ color for the LMC and SMC RRc stars. These differences can be attributed to the significantly different sample sizes used in this work and also in the LMC and SMC \citep{bhardwaj2014}. Furthermore, at $(r-i)_0$ and $(i-z)_0$ colors, temperature fluctuations are expected to have less influence on amplitude variations. Therefore, we do not see a period-amplitude relation for RRc stars in our sample similar to \citet{soszynski2009} for $I$-band data. Another possible cause of this difference can be the selection of $g$-band amplitude as reference for all colors, instead of using amplitude corresponding to the shorter wavelength. For example, we observe that the slope of the $(g-r)_0$ AC relation at maximum light, $-0.093\pm0.054$, is almost identical to the $(V-I)$ AC relation based on the Large Magellanic Cloud RRc stars \citep[$-0.089\pm0.014$,][]{bhardwaj2014}. However, it will be interesting to investigate this further with a greater sample in the future, for example, as provided by the Large Synoptic Survey Telescope (LSST). At minimum light, significant non-zero slopes were found for the AC relations as in previous work \citep{bhardwaj2014}. Further $F$-test results show that the AC relations at minimum light for RRc stars are linear in the $(u-g)_0$ and $(i-z)_0$ colors, and quadratic in other two colors. 

Both \citet{kanbur2005} and \citet{bhardwaj2014} found a significant and flat (or very shallow) AC relations at maximum light and minimum light, respectively, for the RRab stars. Our $t$-test results confirm these earlier works that the AC relations at maximum light exhibit a non-vanishing slope in all four SDSS colors. This is a consequence of the relation between PC and AC, originally developed in \citet{simon1993} and later work (see the Introduction). Some of these relations are better described by a quadratic regression, such as the $(u-g)_0$ and $(i-z)_0$ AC relations for OoI RRab stars and $(g-r)_0$ AC relation for OoII RRab stars. Flat or very shallow AC relations at minimum light are found for OoII RRab stars except the $(g-r)_0$ AC relation. A marginal flat $(u-g)_0$ AC relation at minimum light is also found in OoI RRab stars, but the AC relations in other colors are either linear (for $(r-i)_0$ color) or quadratic (for $(g-r)_0$ and $(i-z)_0$ colors). It is worth mentioning that in $(g-r)_0$ color, the AC relations at minimum light for both OoI and OoII RRab stars are well represented by a quadratic regression, which can also be seen from the right panel in Figure \ref{fig_acmaxmin}.

\section{Discussion and Conclusion}\label{conclu}

The aim of this study is to continue the investigation of empirical PC and AC relations at maximum and minimum light for RR Lyraes in Sloan colors. Such studies can be used to probe the radiation hydrodynamics of the outer envelopes of RR Lyraes (and Cepheids), and potentially be used to estimate the foreground extinction with flat relations. The main differences between this work and previous work, presented in \citet{kanbur2005} and \citet{bhardwaj2014}, include (a) the application of a template light-curve fitting approach to determine the colors at maximum and minimum light instead of using Fourier decomposition technique, (b) separating out the OoI and OoII RRab stars in our sample, which is not done in previous work, and finally (c) the investigation of the relations in four colors based on the SDSS $ugriz$ photometry instead of single color in either $(V-R)_{MACHO}$ or $(V-I)$ color. 

Since we applied the template light-curve fitting approach instead of the usual Fourier decomposition technique as was done in the past \citep{kanbur2005,bhardwaj2014}, we compare the PC relations at maximum and minimum light based on both methods in Figure \ref{fig_vsfourier}. This Figure reveals that the PC relations obtained from the Fourier decomposition technique, using a fixed fitting order, are very similar to those based on template light-curve fitting, including the separation of OoI and OoII RRab stars in the $(g-r)_0$ PC relations at maximum light. This implies that both methods can deliver similar PC relations, and our results are not affected by the choice of fitting methods. However, Figure \ref{fig_vsfourier} also implies that PC relations, especially in the $(u-g)_0$ color, exhibit a larger scatter when using the Fourier decomposition technique. 

\begin{figure*}
  $\begin{array}{cc}
    \includegraphics[angle=0,scale=0.41]{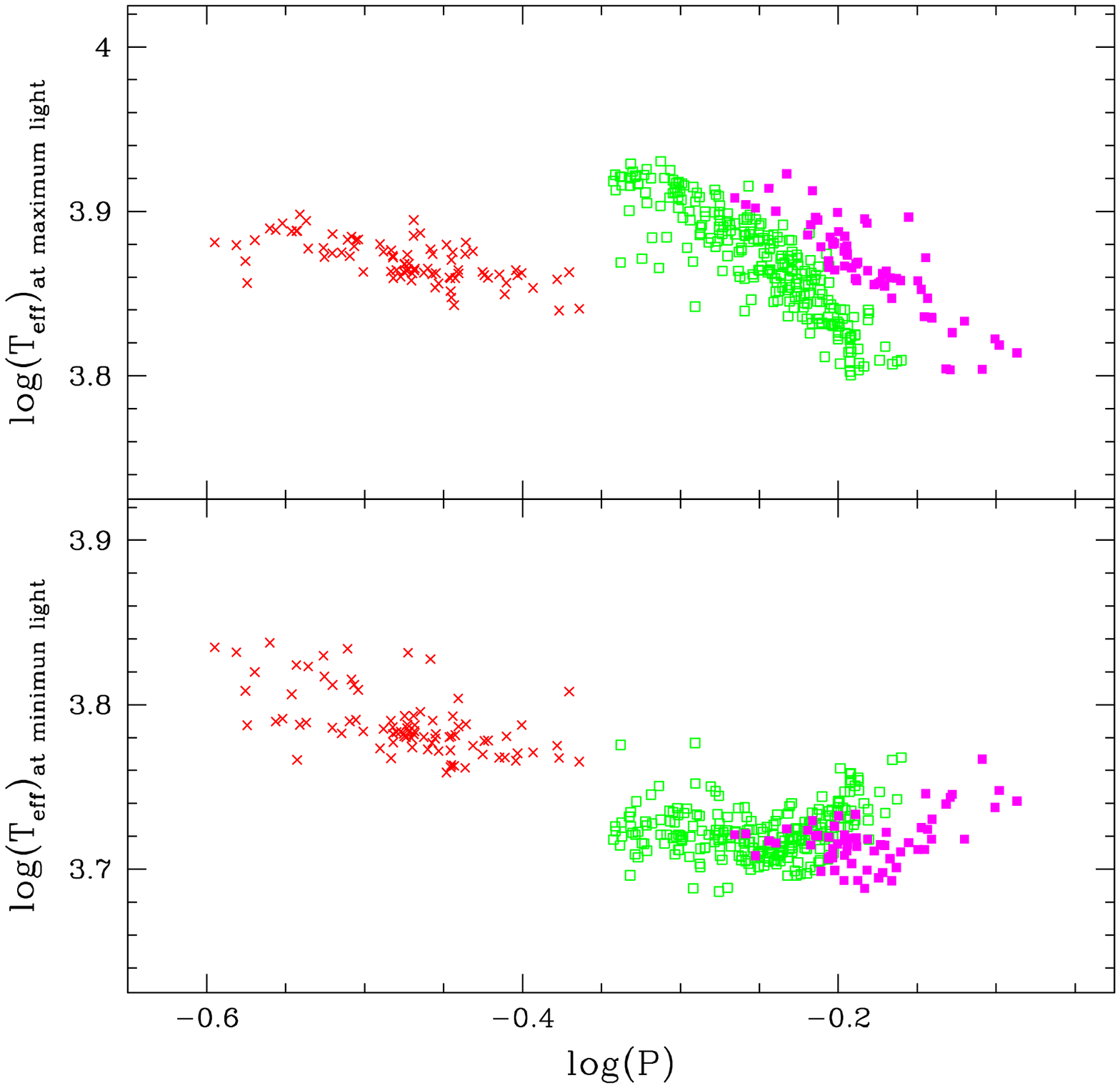} & 
    \includegraphics[angle=0,scale=0.41]{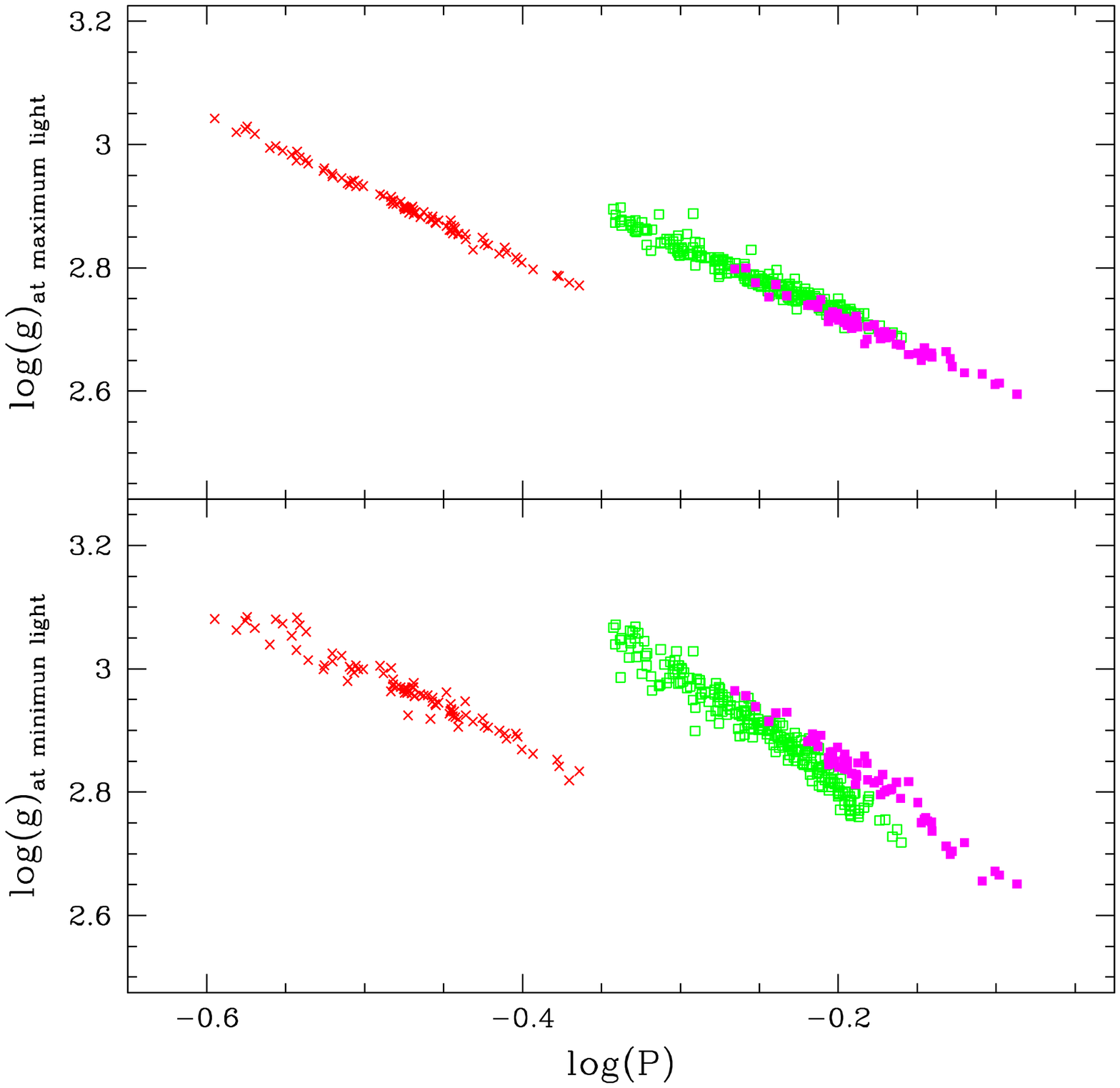} \\
    \includegraphics[angle=0,scale=0.41]{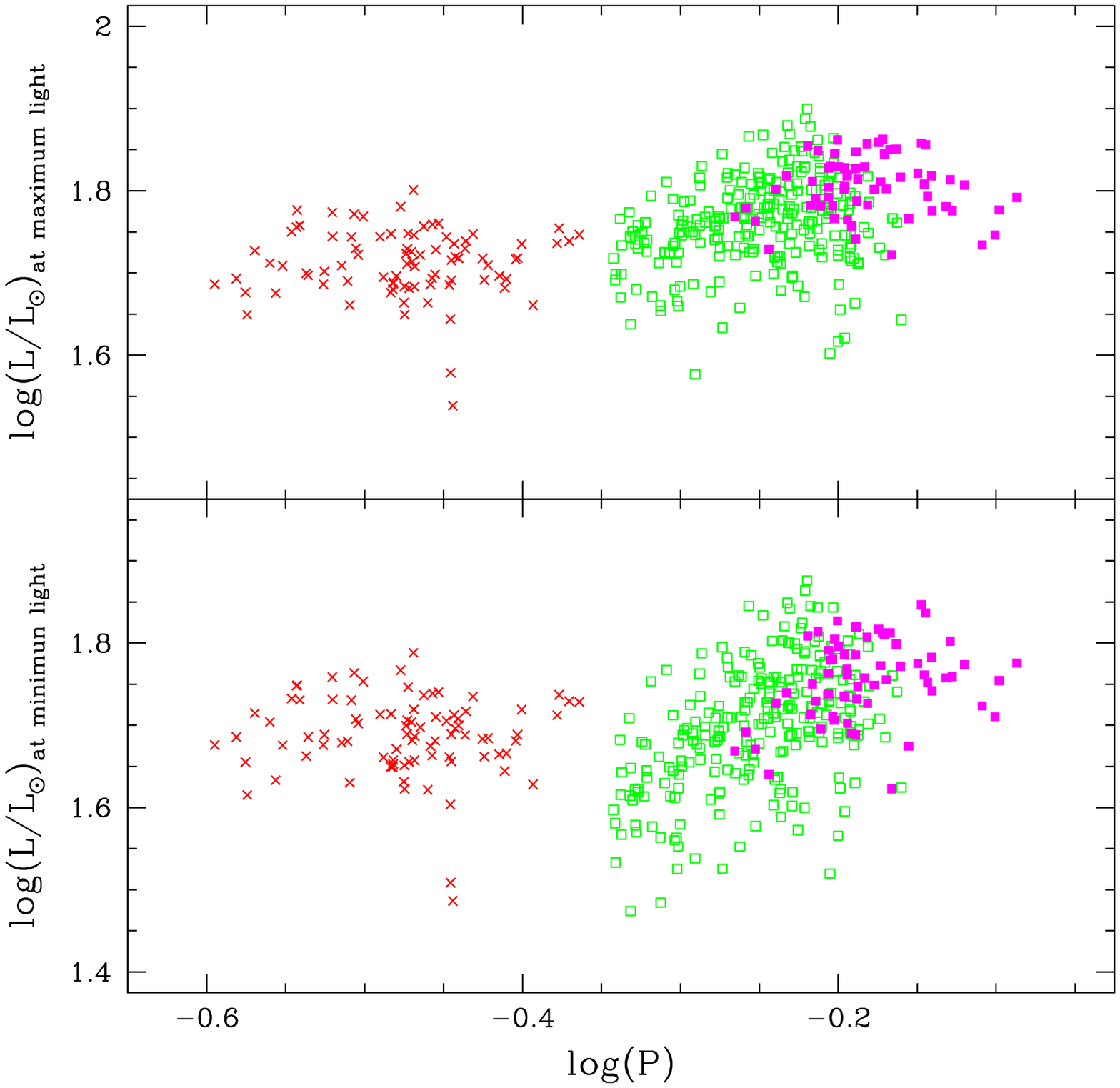} & 
    \includegraphics[angle=0,scale=0.41]{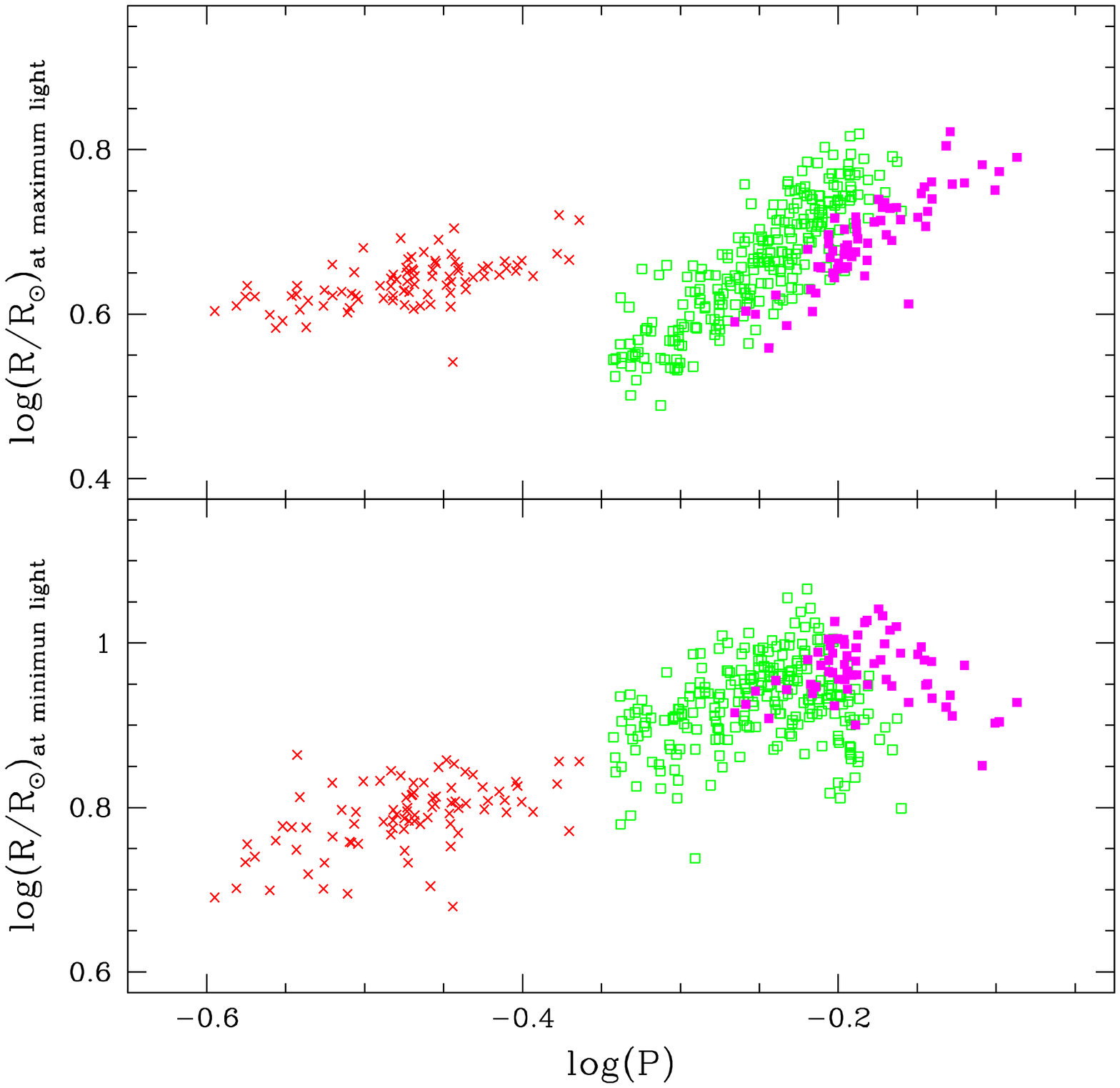} \\    
  \end{array}$ 
  \caption{Period-temperature (upper-left panel), period-gravity (upper-right panel), period-luminosity (lower-left panel), and period-radius (lower-right panel) relations at maximum and minimum light for RR Lyrae in our sample. The effective temperature ($\log~T_{\mathrm{eff}}$), gravity ($\log [g]$), and the luminosity ($\log [L/L_\odot]$) are converted from SDSS colors using the prescription given in \citet{catelan2013}. The adopted conversions involve the pseudo-colors, $C_0 = (u-g)_0 - (g-r)_0$ and/or $m_0 = (g-r)_0 - (r-i)_0$, the $(g-r)_0$ color and $\log P$. Based on the $\log~T_{\mathrm{eff}}$ and $\log (L/L_\odot)$, we calculated the radius, $\log (R/R_\odot)$, by applying the Stefan-Boltzmann law with $\log~T_\odot = 5772$K. For RRc stars, we have fundamentalized the periods \citep[using the relation given in][]{catelan2009} before applying the conversion. Green open squares and magenta filled squares represent the OoI RRab and OoII RRab stars, respectively, while the red crosses are for the RRc stars. \label{fig_convert}}
\end{figure*}

\begin{figure*} 
  \plotone{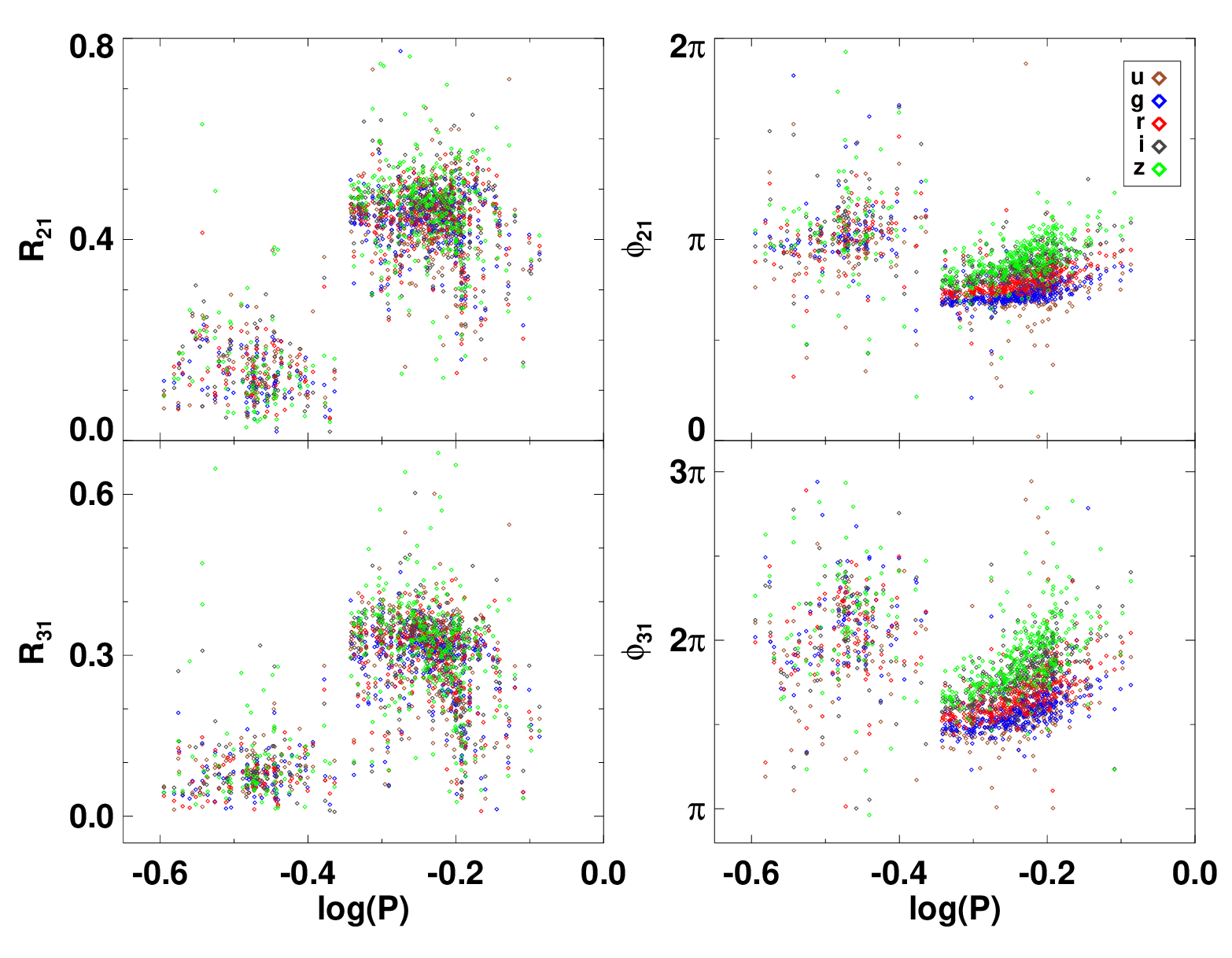}
  \caption{Multi-band lower-order Fourier parameters for the RR Lyrae in our sample. These Fourier parameters are derived from the same Fourier decomposition results as presented in Figure \ref{fig_vsfourier}. The definition of the Fourier parameters ($R_{21}$, $R_{31}$, $\phi_{21}$, and $\phi_{31}$) can be found, for example, in \citet{deb2009} or \citet{bhardwaj2015}. For clarity, we did not separate the OoI RRab stars, OoII RRab stars, and RRc stars in this Figure.} \label{fig_fourier}
\end{figure*}

Our explanation for the PC properties of RR Lyraes relies on the HIF-photosphere interaction \citep{simon1993,kanbur1995,kanbur1996,bhardwaj2014}. As mentioned in the Introduction, the HIF and photosphere are not co-moving during stellar pulsation. Because of this, there are times when the stellar photosphere occurs at the base of the HIF. The times when this occurs varies with the type of star (RR Lyrae or Cepheids), period, and pulsation phase. When the two are not engaged in this way, the temperature of the photosphere and hence the color of the star will be related to the period. For RR Lyraes, the HIF and photosphere are always engaged but only at minimum light, the temperature and density are appropriate such that hydrogen reaches significant levels of ionization at temperatures that are increasingly independent of period. As the star brightens from minimum light, the temperatures that are needed to achieve significant levels of hydrogen ionization become increasingly dependent on density and hence a period-color relation develops. This is consistent with the results described here -- where the slope of the PC relation at minimum light is not strictly zero but is significantly shallower than the slope at maximum light. The fact that the difference in slopes between the PC relations at maximum and minimum light for RRc stars is considerably smaller than that for fundamental mode RRab stars also provides strong support for the HIF-photosphere theory. These RRc stars are typically hotter than RRab stars, and at these temperatures, Saha ionization equilibrium is much more sensitive to the gas density and hence global stellar parameters.

These remarks also apply to AC relations in the following sense. When there is a difference in PC slopes between minimum and maximum light due to the engagement of the HIF-photosphere mentioned earlier, we expect a corresponding difference in the slope of AC relations at maximum/minimum light. As in \citet{simon1993}, we can apply the Stefan-Boltzmann law at maximum and minimum light such that $M_{min} - M_{max} = 10(\log T_{min} - \log T_{max})$, where $M_{min} - M_{max}$ is the amplitude at a given bandpass. Here we neglect the radius terms assuming that temperature fluctuations are more important than radius variations in changing luminosity. Thus if the slope of the PC relation decreases in going from maximum to minimum light, then the slope in an AC plot should decrease in going from minimum to maximum light. Conversely, if the PC slope increases in going from maximum to minimum, then the AC slope should increase in going from maximum to minimum. Table \ref{tab_ac_maxmin} indeed shows this for $(g-r)_0$ and $(r-i)_0$ colors in the case of RRab stars. The evidence is not quite as strong for the $(i-z)_0$ color but in this case one may argue that this color is very much toward the red part of the spectrum so that flux variations due to temperature variations are considerably reduced. In fact, the RRc stars at $(g-r)_0$ and $(r-i)_0$ colors are also consistent with this. 

The behavior of the PC and AC relations for the $(u-g)_0$ color is not consistent with the theory described above, except in the case of OoII RRab stars, which exhibit a shallow PC slope at minimum light, but then the AC relation for these stars has a marginal positive slope. Furthermore the RRc stars have a negative slope in the PC diagram at maximum light. One possibility is that the $u$ and $g$ bands straddle the peak of the Planck blackbody curve at temperatures relevant to RR Lyraes. This perhaps could lead to the opposite-sign slopes seen in Figure \ref{fig_pcmaxmin} for RRc stars. However, our preliminary calculations suggest that $\log(I_u/I_g)$ does not change slope for a range of temperatures between $5000$ and $8000$~K. Thus it may be that bolometric corrections, similar to Figure \ref{fig_vscc2008}, are responsible for this slope change though we leave a detailed discussion of this for future work.

The SDSS colors of the RR Lyrae in Figure \ref{fig_pcmaxmin} can be converted to effective temperature, logarithmic gravity, and bolometric luminosity using the conversions given in \citet{catelan2013}. Figure \ref{fig_convert} presents the converted temperatures, gravities, and luminosities, as well as the radii, as a function of period at maximum and minimum light. For RRab stars at minimum light, we immediately observe that up to $\log P \sim -0.2$, the temperature at the stellar photosphere does not vary much with period,\footnote{Therefore, the slope seen in the $(u-g)_0$ relations must be due to bolometric corrections as discussed in Section 3.1.} and exhibits a change at $\log P \sim -0.20$ for OoI RRab stars and $\log P\sim -0.18$ for OoII RRab stars. For longer periods, the temperature at minimum light for RRab stars increases as the period increases. Therefore, the HIF-photosphere interaction at minimum light seems to occur at $-0.35 \le \log P \le -0.2$ for OoI RRab stars and at $-0.25 \le \log P \le -0.18$ for the OoII RRab stars. In the case of RRc stars, their effective temperatures are hotter than the RRab stars and hence they do not show a flat relation, in agreement with \citet{bhardwaj2014}. At maximum light, the temperature decreases as pulsation period increases for both RRab and RRc stars. Figure \ref{fig_convert} also reveals that the OoI and OoII RRab stars can be well separated in several ways. One example is in the period-temperature and period-radius planes at maximum light. 

Figure \ref{fig_fourier} displays the Fourier parameters for the RR Lyrae light curves in SDSS filters studied here. We see a clear differentiation in the $\phi_{31}$ and $\phi_{21}$ phase parameters with wavelength \citep[similar to Cepheids,][]{bhardwaj2015} and a clear feature at $\log P \sim -0.2$ in the amplitude parameters $R_{21}$ and $R_{31}$ for RRab stars as reported with OGLE-III data. It is striking that this feature occurs at the same period at which changes occur in the plots of $\log (T_{\mathrm{eff}}), \log (L/L_{\odot}), \log (R/R_{\odot})$ against $\log P$ in Figure \ref{fig_convert}. RRc stars are also clearly differentiated on the Fourier parameter - $\log P$ planes. Further data collected from the LSST \citep[for example see][]{oluseyi12} will be very useful in connecting the change in light-curve structural properties, exemplified by the Fourier parameters, to global stellar properties such as PC and AC relations studied here.

In summary, our study reveals that the structural form of PC and AC relations at maximum and minimum lights for RR Lyraes are much more complicated than previously thought \citep{kanbur2005,bhardwaj2014}. They can be expressed as either a flat, a linear, or a quadratic relation in different colors. Furthermore, the PC and AC relations for RRab stars are separated by Oosterhoff types. The empirical results found in this work are certainly worth further theoretical investigations and interpretations, which will be presented in a future paper.

\acknowledgments

We greatly thank the anonymous referee and M. Catelan for providing valuable comments to improve the manuscript. CCN thanks the funding from the Ministry of Science and Technology (Taiwan) under the contract 104-2112-M-008-012-MY3. AB acknowledges the Senior Research Fellowship grant 09/045(1296)/2013-EMR-I from the Human Resource Development Group (HRDG), which is a division of the Council of Scientific and Industrial Research (CSIR), India.

% ===============================================
%               REFERENCE
% ===============================================


\begin{thebibliography}{}

\bibitem[Bhardwaj et al.(2014)]{bhardwaj2014} Bhardwaj, A., Kanbur, S.~M., Singh, H.~P., \& Ngeow, C.-C.\ 2014, \mnras, 445, 2655 

\bibitem[Bhardwaj et al.(2015)]{bhardwaj2015} Bhardwaj, A., Kanbur, S.~M., Singh, H.~P., Macri, L.~M., \& Ngeow, C.-C.\ 2015, \mnras, 447, 3342 

\bibitem[Bontorno et al.(2011)]{bontorno2011} Bontorno, A., Berke, M., Phelps, C., Kanbur, S., \& Ngeow, C.\ 2011, 9th Pacific Rim Conference on Stellar Astrophysics. Proceedings of a conference held at Lijiang, China in 14-20 April 2011. ASP Conference Series, Vol. 451. Edited by S. Qain, K. Leung, L. Zhu, and S. Kwok. San Francisco: Astronomical Society of the Pacific, 2011., p.139

\bibitem[C{\'a}ceres \& Catelan(2008)]{caceres2008} C{\'a}ceres, C., \& Catelan, M.\ 2008, \apjs, 179, 242

\bibitem[Catelan(2009)]{catelan2009} Catelan, M.\ 2009, \apss, 320, 261 

\bibitem[Catelan et al.(2013)]{catelan2013} Catelan, M., Torrealba, G.~I., C{\'a}ceres, C., et al.\ 2013, Advancing the Physics of Cosmic Distances, Proceedings of the International Astronomical Union, IAU Symposium, Volume 289, Edited by Richard de Grijs, and Giuseppe Bono. Cambridge: Cambridge University Press, p.126

\bibitem[Clement(2000)]{clement2000} Clement, C.~M.\ 2000, IAU Colloquium 176: The Impact of Large-Scale Surveys on Pulsating Star Research, ASP Conference Series, Vol. 203, edited by L. Szabados and D. Kurtz, p.266 

\bibitem[Guldenschuh et al.(2005)]{guldenschuh2005} Guldenschuh, K.~A., Layden, A.~C., Wan, Y., et al.\ 2005, \pasp, 117, 721 

\bibitem[Deb \& Singh(2009)]{deb2009} Deb, S., \& Singh, H.~P.\ 2009, \aap, 507, 1729 

\bibitem[Kanbur(1995)]{kanbur1995} Kanbur, S.~M.\ 1995, \aap, 297, L91 

\bibitem[Kanbur \& Phillips(1996)]{kanbur1996} Kanbur, S.~M., \& Phillips, P.~M.\ 1996, \aap, 314, 514 

\bibitem[Kanbur \& Ngeow(2004)]{kanbur2004} Kanbur, S.~M., \& Ngeow, C.-C.\ 2004, \mnras, 350, 962 

\bibitem[Kanbur et al.(2004)]{kanbur2004a} Kanbur, S.~M., Ngeow, C.-C., \& Buchler, J.~R.\ 2004, \mnras, 354, 212 

\bibitem[Kanbur \& Fernando(2005)]{kanbur2005} Kanbur, S.~M., \& Fernando, I.\ 2005, \mnras, 359, L15 

\bibitem[Kanbur \& Ngeow(2006)]{kanbur2006} Kanbur, S.~M., \& Ngeow, C.-C.\ 2006, \mnras, 369, 705 

\bibitem[Kanbur et al.(2007)]{kanbur2007} Kanbur, S.~M., Ngeow, C.-C., \& Feiden, G.\ 2007, \mnras, 380, 819 

\bibitem[Kanbur et al.(2010)]{kanbur2010} Kanbur, S.~M., Marconi, M., Ngeow, C., et al.\ 2010, \mnras, 408, 695 

\bibitem[Kunder et al.(2010)]{kunder2010} Kunder, A., Chaboyer, B., \& Layden, A.\ 2010, \aj, 139, 415 

\bibitem[Layden et al.(2013)]{layden2013} Layden, A., Anderson, T., \& Husband, P.\ 2013, arXiv:1310.0549 

\bibitem[Nelder \& Mead(1965)]{nelder1965} Nelder, J.~A. \& Mead, R.\ 1965, Computer Journal, 7, 308

\bibitem[Oluseyi et al.(2012)]{oluseyi12} Oluseyi, H.~M., Becker, A.~C., Culliton, C., et al.\ 2012, \aj, 144, 9 

\bibitem[Sesar et al.(2010)]{sesar2010} Sesar, B., Ivezi{\'c}, {\v Z}., Grammer, S.~H., et al.\ 2010, \apj, 708, 717 

\bibitem[Simon et al.(1993)]{simon1993} Simon, N.~R., Kanbur, S.~M., \& Mihalas, D.\ 1993, \apj, 414, 310 

\bibitem[Smith et al.(2011)]{smith2011} Smith, H.~A., Catelan, M., \& Kuehn, C.\ 2011, RR Lyrae Stars, Metal-Poor Stars, and the Galaxy, Astrophysics Series, Vol. 5, edited by Andrew McWilliam. Pasadena, CA: The Observatories of the Carnegie Institution of Washington, p.17

\bibitem[Soszy{\'n}ski et al.(2009)]{soszynski2009} Soszy{\'n}ski, I., Udalski, A., Szyma{\'n}ski, M.~K., et al.\ 2009, \actaa, 59, 1 

\bibitem[Sturch(1966)]{struch1966} Sturch, C.\ 1966, \apj, 143, 774 

\end{thebibliography}
\end{document}